\documentclass[fleqn,usenatbib,useAMS]{mnras}

\usepackage{graphicx}	% Including figure files
\usepackage{amsmath}	% Advanced maths commands
\usepackage{amssymb}	% Extra maths symbols
\usepackage{multicol}        % Multi-column entries in tables
\usepackage{bm}		% Bold maths symbols, including upright Greek
\usepackage{pdflscape}	% Landscape pages
\usepackage{multirow}
\usepackage{listings}

\def\nstepscut{{\it nsteps\_cut}}
\def\nsteps{{\it nsteps}}
\def\nwalkers{{\it nwalkers}}
\def\sourcefile{{\it source\_file}}
\def\modelfile{{\it model\_file}}
\def\perc{{\it percentiles}}
\def\z{{\it redshift}}

\def\mrmoose{Mr-Moose}

\def\herschel{{\it Herschel}}

\def\mum{\,$\mu$m}

\def\spitzer{{\it Spitzer}}
\usepackage[T1]{fontenc}
\usepackage{ae,aecompl}

\usepackage{newtxtext,newtxmath}
\title[Mr-Moose]{\textit{Mr-Moose}: An advanced SED-fitting tool for heterogeneous multi-wavelength datasets}

\author[G. Drouart, T. Falkendal]{G. Drouart$^{1}$\thanks{Contact e-mail: \href{mailto:guillaume.drouart@curtin.edu.au}{guillaume.drouart@curtin.edu.au}} \& T. Falkendal$^{2,3}\thanks{Contact e-mail: \href{mailto:tfalkend@eso.org}{tfalkend@eso.org}}$
\\
$^{1}$International Centre for Radio Astronomy Research, Curtin University, 1 Turner Avenue, Bentley, Western Australia 6102, Australia \\
$^{2}$European Southern Observatory, Karl Schwarzschild Stra\ss e 2, 85748 Garching bei M\"unchen, Germany \\
$^{3}$Institut d'Astrophysique de Paris, 98bis boulevard Arago, 75014 Paris, France}

\date{28 March 2018}

\pubyear{2018}

\begin{document}
\label{firstpage}
\pagerange{\pageref{firstpage}--\pageref{lastpage}}
\maketitle

% Abstract of the paper
\begin{abstract}

We present the public release of \mrmoose, a fitting procedure that is able to perform multi-wavelength and multi-object spectral energy distribution (SED) fitting in a Bayesian framework. This procedure is able to handle a large variety of cases, from an isolated source to blended multi-component sources from an heterogeneous dataset (i.e. a range of observation sensitivities and spectral/spatial resolutions). Furthermore, \mrmoose\ handles upper-limits during the fitting process in a continuous way allowing models to be gradually less probable as upper limits are approached. The aim is to propose a simple-to-use, yet highly-versatile fitting tool fro handling increasing source complexity when combining multi-wavelength datasets with fully customisable filter/model databases. The complete control of the user is one advantage, which avoids the traditional problems related to the ``black box'' effect, where parameter or model tunings are impossible and can lead to overfitting and/or over-interpretation of the results. Also, while a basic knowledge of Python and statistics is required, the code aims to be sufficiently user-friendly for non-experts. We demonstrate the procedure on three cases: two artificially-generated datasets and a previous result from the literature. In particular, the most complex case (inspired by a real source, combining {\it Herschel}, ALMA and VLA data) in the context of extragalactic SED fitting, makes \mrmoose\ a particularly-attractive SED fitting tool when dealing with partially blended sources, without the need for data deconvolution.

\end{abstract}

% Select between one and six entries from the list of approved keywords.
% Don't make up new ones.

%%%%%%%%%%%%%%%%%%%%%%%%%%%%%%%%%%%%%%%%%%%%%%%%%%

%%%%%%%%%%%%%%%%% BODY OF PAPER %%%%%%%%%%%%%%%%%%
\section{Introduction}

Confronting observations and models of our Universe is the very core of any fitting procedure. However, the fitting process 
faces an increasing number of challenges to adapt and compare the wide variety of models and observations. From the modelling side,
generations of large libraries of highly non-linear physical processes allow for a more accurate description of the complexity of
the physical processes at work in the sources. From the observer's side, new instruments and facilities continue to push further resolution --- both spectrally and spatially --- and sensitivity over the electromagnetic spectrum. The principle of fitting appears at the frontiers of these two sides and interfacing these two aspects optimally is crucial. How to obtain maximal information from the data (which can vary a lot on quality and quantity) and compare it to the most optimal models available? By optimal here, we define the model which will represent data with the most fidelity or complexity without going in the regime of "overfitting", where degeneracies between parameters dominate, and therefore, no meaningful constraints can be extracted from the dataset (usually keeping in mind the Ockham razor principle). This optimal regime can be quantified thanks to the Bayes theorem, where likelihood of different models can be compared to find the relative likelihood probability of one model compared to another. The particular case of spectral energy distribution fitting is an old problem and can be traced as far as the 1960s \citep[e.g.][]{Tinsley1968}. This technique has seen a considerable development starting at the end of the twentieth century and is now widely used in most fields of Astrophysics: from stars in our Galaxy to very distant galaxies during the epoch of reionisation. 

\newpage
As data throughout the electromagnetic spectrum became readily available, more and more complex models were developed, such as PEGASE \citep{Fioc1997}, \cite{Maraston1998}, {\it Starburst99} \citep{Leitherer1999}, \cite{Bruzual2003},  MAPPINGS III \citep{Groves2004} to predict galaxy emission and \cite{Pier1992, Honig2006, Fritz2006, Nenkova2008a, Stalevski2012, Siebenmorgen2015} for AGN emission to name the most widely used. Note that these codes do not contain any fitting routines, their main intent being to provide the user with a vast library of templates exploring a large, often non-linear parameter space.  As a result, SED fitting techniques remained primitive, mainly using minimising algorithms and regression such as least-square or chi-square minimisation. Since 2000, we have seen a rise of tools implementing more sophisticated statistical frameworks in order to explore degeneracies in the increasing complexity of the SED fitting. It is important here to distinguish the two complementary aspects of SED fitting: the models used to predict the SED given a set of parameters and assumptions \citep[][for a recent review]{Conroy2013} and the algorithm used to estimate the parameters on data \citep[the fitting itself, see][for a review on the different available algorithms]{Sharma2017}. During the last decade, these two aspects were often combined to provide users off-the-shelf SED fitting procedures with a varying wavelength range: Magphys \citep{daCunha2008} for galaxy evolution in the UV-FIR range, BayesClumpy \citep{AsensioRamos2009} for AGN torus fitting in the UV-FIR range, CIGALE \citep{Noll2009} for galaxy evolution with AGN implementation in the UV-FIR range, BRATS \citep{Harwood2013} specialised for multi-radio frequency fitting in resolved sources, SEDfit \citep{Sawicki2012} for galaxy evolution in optical/NIR implementing spatially resolved sources, SEDeblend \citep{MacKenzie2016} for SED fitting in the context of lensed source with FIR observations, AGNFitter \citep{CalistroRivera2016} focusing on the AGN component in the UV-FIR range, ... to name the most widely used in extragalactic astronomy\footnote{see also \url{http://www.sedfitting.org/Fitting.html} for a more exhaustive listing.}.

In an extragalactic context, most of these codes such as Magphys, CIGALE, BayesClumpy or AGNFitter, rely on a relatively strong assumption: all the emission comes from a single unresolved source and cross-identification at different wavelength is trivial. This simplification, while being necessary to enable the interpretation of extended multi-wavelength coverage (typically from UV to FIR) and to make use of the numerous source catalogues available in literature, is also one of their most important shortcomings in the advent of high resolution imaging across the electromagnetic spectrum. Indeed, as the wavelength range increases, resolution and sensitivity change drastically, and the assumption of the single source is no longer valid. Also, the other codes such as BRATS, SEDeblend or SEDfit are specialised in relatively short wavelength range and/or very specific applications: BRATS for spectral index mapping in radio, SEDeblend for lensed sources in the FIR domain and SEDfit in the optical/NIR domain for resolved galaxies. None of the aforementioned codes present the possibility to fit partially resolved sources, from NIR to radio. 

To cite a more specific example as presented in \citet{Gullberg2016}: fitting simultaneously \spitzer\citep[3-160\mum][]{Werner2004}, \herschel\citep[70-500\mum][]{Pilbratt2010}, ALMA(100-700\,GHz) and VLA data for distant, partially resolved galaxies, illustrates exactly this point. One one hand, the resolution of \herschel\ does not allow to properly disentangle different spatial components (blending) but provides five data-points in spectral space, particularly covering the thermal dust IR peak of the SED. One the other hand, ALMA is particularly suited to answer to the question of the spatial location, at the cost of smaller spectral coverage. Also the heating sources of the dust emitting at 70\mum\ and 3\,mm is different (AGN or young stars) but is linked through the total energy emitted and their location. The same type of reasoning can be applied to the AGN radio emission ($\nu$$<$100\,GHz, from the VLA for instance), often unresolved at low frequency and presenting multi-components at higher frequencies (core, jets, hot-spots and lobes). In the case of radio galaxies, these last two aspects are even co-existing for a single source. Therefore, as these observations are complementary, each providing valuable constraints spectrally and spatially on different overlapping components (some with non-detections), why not combining all information together? This is where \mrmoose\ intervenes, providing a framework to simultaneously combine spatial and spectral information into a single fitting routine. The paper is organised as follows: we explain the methodology and structure of the code as well as the description of the input and output files in \S~\ref{sec:design}. \S~\ref{sec:examples} presents three examples of application of \mrmoose. In \S~\ref{sec:limitations}, we discuss the limitations and future developments and finally conclude in \S~\ref{sec:conclusion}.

\section{Design and structure of \mrmoose}
\label{sec:design}

\mrmoose\ stands for {\b M}ulti-{\b r}esolution and {\b m}ulti-{\b o}bject/{\b o}rigin {\b s}pectral {\b e}nergy distribution fitting procedure. The philosophy when designing \mrmoose\ was to provide a tool for SED fitting, sufficiently simple to be used by non-SED fitting experts, yet sufficiently versatile to handle as many cases as possible, particularly in the context of multi-wavelength samples. The main drivers for the design are:
\begin{enumerate}
\item to handle data with a large span of resolutions to combine interferometer/single-dish observations,
\item to deal with upper limits consistently, allowing a fit with gradually decreasing probability when approaching an upper-limit rather than an arbitrary cut at 3$\sigma$,
\item to treat the fitting in a bayesian framework to provide asymmetric uncertainties on parameters and to identify easily degeneracies as well as comparing different model combinations,
\item to allow a wide variety of models as input, analytic models for this first release and/or libraries for non-linear models in future updates.
\end{enumerate}

\begin{figure*}
\centering
\includegraphics[width=\textwidth]{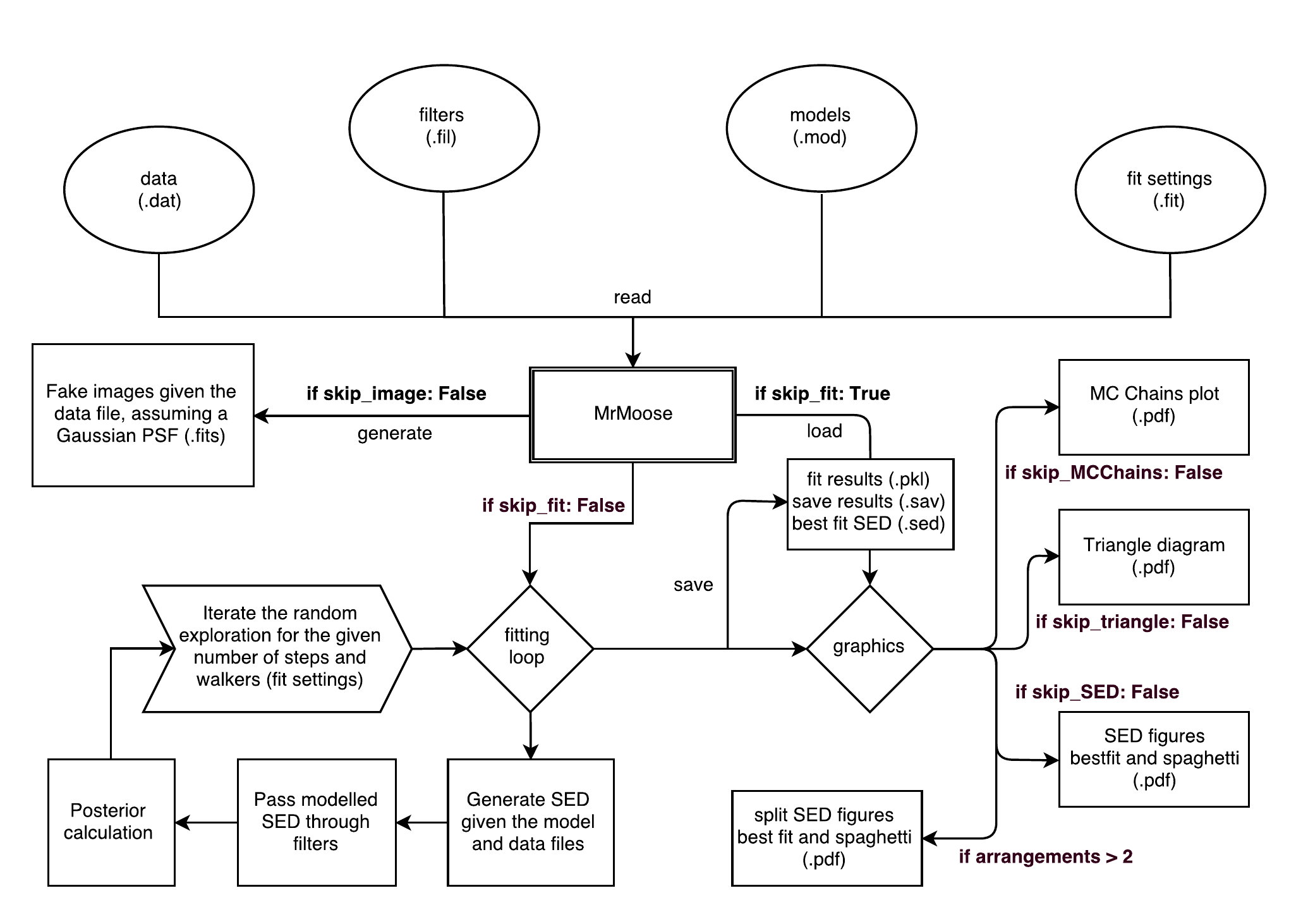}
\caption{Flow chart of \mrmoose.}
\label{fig:flowchart}
\end{figure*}

\mrmoose\ is written in python 2.7, developed in PyCharm Community Edition and complies to the PEP8 writing standard. It relies as much as possible on already existing packages (see full list in Appendix) to further increase robustness. The code was designed to be user friendly but not user opaque. A large (and necessary) control of the data, model and fit setting inputs is allowed to test the reliability of the output and therefore avoid the ``black-box'' effect. However, the code aims to be sufficiently simple to be used with a minimal statistical and computational knowledge. We first present the core of the fitting procedure, the maximum likelihood estimation and review the inputs and outputs files. Table \ref{tab:extension} summarises the different file extensions used in \mrmoose. We also report a simplified flowchart of the code structure in Figure \ref{fig:flowchart} and refer the user to a more detailed flowchart in the GitHub repository\footnote{\url{https://github.com/gdrouart/MrMoose}}.

\begin{table*} \centering
\caption{Summary of the convention for the file extensions, as used in \mrmoose}
\label{tab:extension}
\begin{tabular}{lll}
\hline
Extension & Use & Description \\
\hline \hline
.fit & input & file containing the settings for the fit. \\
.dat & input & file containing the data. \\
.mod & input & model file containing the function, parameters and their considered range \\
.fil & input & filter files, containing the transmission curves \\
.sav & output & file containing the information from the fit, saved for future use, and storing the settings for the fit \\
.sed & output & file containing the flux per frequency (in Hz) of the best fit for all components of the SED in unit of erg s$^{-1}$cm$^{-2}$Hz$^{-1}$ \\
.pdf & output & output files containing the plots \\
.fits & output & modelisation of the source(s) in each filter from the .dat file \\
.py & ... & \mrmoose\ files containing the procedures. \\
\hline
\end{tabular}
\end{table*}

\subsection{The maximum likelihood estimation in \mrmoose}

The code working in a bayesian framework, likelihood calculation is central in identifying the most representative models to the data and providing marginalised probability density functions for each parameter in order to estimate uncertainties. Following the Bayes theorem written in terms relevant for SED fitting, the probability of a model M given the data D can be written as:

\begin{equation}
P(M|D) = \frac{P(M) * P(D|M)}{P(D)},
\end{equation}

where P(M|D) is the posterior distribution (the probability of M {\it after} observing D), P(M) is the prior probability (the probability of M {\it before} observing D), P(D|M) is the likelihood (the probability of observing D given M), and P(D) is the model evidence (a measure of how well the model M predicts the data D and can be considered as a normalisation factor to the problem, often non-trivial to estimate). Therefore, $P(M|D)$ is the final parameter distribution (that we are looking for) and $P(D|M)$ is likelihood to be determined and to maximise (or minimise given our estimator, a slightly modified version of the goodness-of-fit). This maximum likelihood (abbreviated P onwards) corresponds to the product of all probabilities of the model given the data and can be written as:
\begin{equation}
P \propto \prod_{i}P_{i}\prod_{j}P_{j},
\label{eq:gen}
\end{equation}
with $i$ the detections and $j$ the upper limits. We stress that the prior on the model parameters (set to uniform, corresponding to an uninformative prior in \mrmoose\ version 1). Each individual probability distribution, respectively for detection and upper limit (following a normal distribution), are defined as follows: 
\begin{equation}
P_{i} \propto {\rm exp} \left[-\frac{1}{2}\left( \frac{S_{i}^{data} - S_{i}^{model}}{\sigma_i} \right)^2 \right]\Delta S_{i},
\label{eq:det}
\end{equation}
\begin{equation}
\label{eq:ul}
P_{j} \propto \int_{-\infty}^{S_{lim,j}^{data}} {\rm exp} \left[-\frac{1}{2}\left( \frac{S_{j} - S_{j}^{model}}{\sigma_j} \right)^2 \right]dS_{j},
\end{equation}
where $S_{i}^{data}$ is the measured flux and $\sigma_i$, its standard deviation, $S_{i}^{model}$ is the predicted flux, all in filter $i$ for a detection and $j$ for a non-detection. Note the $\Delta S_{i}$ as being the data offset from the true value of $S_{i}^{data}$, infinitesimally small in the case of detection, and $dS_{j}$ referring to the integrated flux probability ($S_{j}$) to the detection threshold $S_{lim,j}^{data}$ \citep[see Appendix 1, their Fig. 6, ][for a graphical representation]{Sawicki2012}. Taking the logarithm of Eq. \ref{eq:gen} and injecting Eq. \ref{eq:det} and \ref{eq:ul}, we obtain:
\begin{equation}
\label{eq:prob}
\begin{split}
P \propto & -\frac{1}{2} \sum_{i} \left( \frac{S^{data}_{i}-S^{model}_{i}}{\sigma_{i}} \right)^2 + \sum_{i} {\rm ln} \Delta S_{i} \\ &+ \sum_{j} {\rm ln} \int_{-\infty}^{S_{lim,j}^{data}} {\rm exp} \left[-\frac{1}{2}\left( \frac{S_{j} - S_{j}^{model}}{\sigma_j} \right)^2 \right]dS_{j},
\end{split}
\end{equation}
The second term of the right-hand side of the equation being constant, maximising the probability, $P$, corresponds to minimising the following estimator:
\begin{equation}
\label{eq:chi2}
\begin{split}
\chi^2 = & \sum_{i} \left( \frac{S^{data}_{i}-S^{model}_{i}}{\sigma_i} \right)^2 \\
& -2 \sum_{j} ln \left\{ \sqrt{\frac{\pi}{2}} \sigma_{j} \left[1 + {\rm erf}\left(\frac{S^{data}_{lim,j}-S^{model}_{j}}{\sqrt{2}\sigma_{j}}\right)\right] \right\}.
\end{split}
\end{equation}

We apply a change of expression of the third term of Eq.~\ref{eq:prob} making use of the error function, {\it erf}, for computational convenience (see Appendix A of \citet{Sawicki2012} for details). In practice, the error function tends to infinity quickly but gradually after a threshold point (the upper limit) calculated by the user as the local noise level in the original image, in the corresponding filter (the same image at a given filter/frequency can provide a combination of detections and upper limits if the location of components are known). This $\chi^2$ is only slightly different when including the upper limits into the calculation. We note that in the case when no upper limit is provided, the second term is null, and therefore the equation corresponds to the classical $\chi^2$ definition. In order to explore the parameter space and minimises the $\chi^2$ (equivalent to maximising the posterior probability), we make use of the {\tt emcee} python package \citep{emcee2013}, allowing a choice of different samplers to probe the parameter space. 
%In a nutshell, this algorithm simultaneously allows a large number of Markov chains (walkers) to explore a wide parameter space, enabling fast convergence even in more complex parameter space. 

The {\tt emcee} package follows the classic Markov chain Monte Carlo (MCMC) approach, consisting of randomly moving walkers exploring the parameter space. The main goal is to sample of the posterior probability density function of the parameters with a sufficiently high number of independent realisations. This is the best compromise between a brute force approach where each solution of the parameter space is calculated for a given grid and a standard least-square minimisation where only the best solution is provided (which can be a local minima). This approach also presents the advantage to estimate uncertainties on the model parameters, even with non-normal distributions thanks to the exploration the parameter space, but without the computing cost (increasing exponentially with the number of parameters). To explore and reach convergence on the best solution, the main idea of a MCMC approach is to set a walker in the parameter space, moving randomly with a given step size. For each new position, the likelihood function is evaluated, and the step is accepted if the probability is higher than the previous position. Hence the walker moves slowly towards the best solution which maximise the likelihood function. Several algorithms are available, the most famous being the Metropolis-Hasting (M-H) algorithm. However, this algorithm can fail to converge quickly when the convergence parameters are not optimally tuned and the parameter space presents strong non-linear behaviour (in particular if the step length is not optimised for the problem). New algorithms were developed to answer these caveats, reducing significantly the converging time. In particular, {\tt emcee} uses affine-invariant transformations \citep{emcee2013, Goodman2010} to sample the parameter space, using a stretch move based on the other walker positions. This presents the advantages of being less sensitive to strongly correlated parameters allowing a minimum loss of performance and therefore a quicker convergence. Also, this algorithm enables two valuable simplifications compared to the standard M-H: (i) the performance of the code is basically relying on two parameters: $a$, a gain parameter and $n$, the number of walkers and (ii) the possible parallelisation, splitting the chains into complementary sub-ensembles to predict iteratively the next walker moves, enabling even larger performance gains for specific problems. We refer the reader to \citet{emcee2013} for more details on the algorithm.

\subsection{Inputs}

\mrmoose\ requires four input files: a data-file, a model-file, a setting-file, and the filter database each containing the necessary information to perform the fit (see Table \ref{tab:extension}).

\subsubsection{Data file}
The data-file requires the most attention in its design as it will set up the number of arrangements (the various data/model combination). The \mrmoose\ procedure doesn't check for inconsistency in the data, and is the responsibility of the user to provide reliable or relevant photometric points (well calibrated and corrected for any photometric effects). To help with this step, corresponding images will be generated for each reported filter in the data file (see \S~\ref{sec:out_fits}). Multi-wavelength SED fitting is particularly challenging in this regard, therefore specific care should be taken in order to have consistent data throughout the fitting range (blending, filtering of scale/flux by interferometers, instruments corrections, aperture effects, absolute calibration uncertainties) by adding extra models/arrangements if necessary. However, we stress that starting with a simple case and gradually complexifying the fitting is usually a rewarding approach. Table~\ref{tab:dat} summarises the data-file organisation.

\begin{table*} \centering
\caption{Structure of the .dat file. We provide here an example}
\label{tab:dat}
\begin{tabular}{l l p{12cm}}
\hline
Name of columns & Type & Purposes/Notes \\
\hline \hline
{\it Filter} & string & filter name. It should match exactly the name of a filter in the database (case sensitive), without the extension (***.fil), necessary to calculate the model flux during the fitting (\S~\ref{sec:filter_database}) \\
{\it lambda0} & float & central wavelength of the filter \\
{\it RA} & string & right ascension of the source in the 00h00m00.0s format - used only in the .fits generation \\
{\it Dec} & string & declination of the source in the 00d00m00.0s format - used only in the .fits generation \\
{\it resolution} & float & resolution of the instrument in arcsec - used only in the .fits generation \\
{\it det\_type} & string & indicate if the reported value is a detection or an upper limit \\
{\it flux} & float & observed flux in the given filter \\
{\it flux\_error} & float & uncertainty on the flux for the given filter \\
{\it arrangement} & integer & the number of the arrangement \\
{\it component} & string & the name of the arrangement (will be used for the spilt plot) \\
{\it component\_number} & list of integer & the list of models to fit for this data point (separated with a semi-column) \\
\hline
\end{tabular}
\end{table*}

\subsubsection{Model file}
\label{sec:models}

The model-file contains the information of the parameters from the combination of the models to fit on the data. Table~\ref{tab:mod} shows the format of the model file. The function name is referring to the function to be called during the fit and therefore should match exactly the name of the user defined function (case sensitive). The function is to be defined in the {\tt models.py} file. Ideally, it should be sufficiently well designed to avoid a drop in execution time as each model is executed at each step for each walker by avoiding as much as possible loops, conditions or function calls. The code already provides some functions, developed and tested for specific applications and examples \citep[][]{Gilli2014,Falkendal2017, Drouart2017}. We provide here a description of the models provided with the release of \mrmoose:
\begin{itemize}
\item BB\_law: a black body model directly called from the {\tt astropy} package \citep{astropy2013}
\begin{equation}
S_{\nu} = N \frac{2h\nu^3/c^{2}}{{\rm exp}(h\nu/kT)-1},
\end{equation}
where $S_{\nu}$ is the flux in erg s$^{-1}$ cm$^{-2}$ Hz$^{-1}$, $N$ the normalisation to the data (relatively complex term containing the scaling to the data and the solid angle covered by the source with the small angular size source simplification, depending on each instrument for unresolved sources see\footnote{\url{https://casper.berkeley.edu/astrobaki/index.php/Single_Dish_Basics}} for more details), $\nu$ the frequency in Hz, $h$ the Planck constant, $k$ the Boltzmann constant, $T$ the temperature in K, $c$ the speed of light.
\item AGN\_law: an empirical AGN model designed to fit the IR AGN component
\begin{equation}
S_{\nu} = N \left(\frac{\nu_{cut}}{\nu_{obs}}\right)^{-\alpha} {\rm exp} \left(-\frac{\nu_{cut}}{\nu_{obs}}\right),
\end{equation}
where $\nu_{cut}$ the cutoff frequency, $\alpha$ the spectra index, $\nu_{obs}$ the observed frequency in Hz
\item sync\_law: a single power law model, common form to represent synchrotron emission in radio and is defined as:
\begin{equation}
S_{\nu} = N\nu^{\alpha},
\end{equation}
\item a modified black body function as used in \cite{Gilli2014} for example \#3 (see \S~\ref{sec:ex3}),
\begin{equation}
S_{\nu} = N B_{\nu}(T)\left(1- {\rm exp}\left[-\left(\frac{\nu}{\nu_0}\right)^{\beta}\right]\right),
\end{equation}
where $B_{\nu}$ is the blackbody function defined previously (as BB\_law), $\beta$ the dust emissivity and $\nu_0$ the frequency where the dust emission enters the optically thick regime ($\tau=(\nu/\nu_0)^\beta=$1).
\end{itemize}

Each model exists in two versions, the standard where an array of frequency, an array with the parameter and the redshift value is provided and a version with redshift as a free parameter by simply adding the \_z at the end of the function (see examples in the \mrmoose\ repository) and by changing the {\tt all\_same\_redshift} and providing the relevant redshift array (one redshift value per component). Note that when the redshift is a free parameter, the inputs are slightly different with only an array of frequency and an array for the parameters. We refer the user to the examples available in the repository. The library of models will increase in future updates depending on the need, we refer to the user manual for an up-to-date library of available models. We remind that the package {\tt astropy} \citep{astropy2013} provides a large library of the most commonly used function in astronomy. Also, we suggest to design the model with parameters spanning a large range in log-scale (such as normalisation factors, as done in the examples in this paper), making sure the corresponding function is making the adequate transformation to calculate flux. The initial position of the walkers are set following the {\tt emcee} documentation: a Gaussian ball of walkers, centred on the median of the parameter interval. This has important consequences and should be taken into account when defining this interval (see \S~\ref{sec:limitations}).

\begin{table} \centering
\caption{Structure of the .mod file. It consists of series of blocs, to be repeated as many time as necessary to add the models to be fitted on the data. We present an example in \S~\ref{sec:ex1} and \S~\ref{sec:ex2}. It is to be noted that the name of the parameters should be expressed in latex syntax if the user requires the parameter to be written in latex format in the plot (particularly the triangle plot). }
\label{tab:mod}
\begin{tabular}{l l l}
\hline
Name &  &  \\
\hline \hline
\# & & \\
{\it function\_name} & number of parameter ($n$) & \\
{first parameter} & min value & max value \\
{second parameter} & min value & max value \\
\vdots & \vdots & \vdots \\
{n parameter} & min value & max value \\
\# & & \\
\hline
\end{tabular}
\end{table}

\subsubsection{Setting file}

The setting file (see full structure in Table \ref{tab:setting_file}) contains the parameters for \mrmoose\ to perform the fit such as the number of walkers (\nwalkers), the number of steps (\nsteps), the limit (\nstepscut) at which the chains should be used to plot the results (the limit from which convergence of the fit is reached). We provide examples with typical numbers. For more information, we invite the user to read the recommendations from the {\tt emcee} documentation. In a nutshell: the more walkers and steps, the better. However a balance has to be found to keep the code executable on a machine. From experience, one effective strategy consists to run some first blind tests to have a feeling of the likely requirements for the number of walkers and the number of steps. For relatively simple case, we advice to start with 200 walkers and run for 500 steps, with a cut at 400 (\nwalkers=200, \nsteps=500 and \nstepscut=400). This first run will probe the parameter space with 20 000 points ((\nsteps-\nstepscut) $\times$ \nwalkers) and gives an indication of how many steps might be required to reach convergence. For a larger the number of parameters, a larger number of walkers and steps is required to explore more thoroughly the parameter space (see example 1 versus example 2). 

\subsubsection{Filter database}
\label{sec:filter_database}

The filter database is essential in the fitting routine as allowing to predict the flux of a given combination of components at a given wavelength. For each step, the average flux of the model is calculated through the transmission curve of the filter given in the data file. This transmission curve represents the response of a telescope for a constant incident flux (in the frequency space) for a given frequency range. The convolution of the incident flux with the telescope response is important as any large variation in the SED within a spectral window can bias the estimated flux, for instance in the case of an emission line or a strong gradient within the averaged frequency range. This effect is particularly strong in optical and NIR where the telescope response varies significantly from filter to filter. During the fitting, we therefore calculate for each filter, the corresponding telescope response convoluted with the modelled source with the following equation:
\begin{equation}
\label{eq:filter}
S_{i}=\frac{\int_{\nu_{min}}^{\nu_{max}} S_{\nu}^{model}t_{\nu} d\nu}{\int_{\nu_{min}}^{\nu_{max}} t_{\nu} d\nu}
\end{equation}
where $S_{i}$ the averaged flux for a given filter $i$, $\nu$ the frequency, $S_{\nu}^{model}$ the flux of the model and $t_{\nu}$ the transmission curve. Each of the filter used in the data file must be present in the database, or the code will fail to execute. Addition of new filters is easy, a simple text file with two columns, frequency and corresponding transmission can be added in the filters folder. For optical and NIR, given the large amount of filters used in astronomy, we point the reader to Asiago Database on Photometric Systems\footnote{\url{http://ulisse.pd.astro.it/Astro/ADPS/}} to add the relevant filters for a specific analysis \citep{Fiorucci2003}. For other wavelength, we refer the user to each instrument user manual to provide the relevant transmission curve. In the case of radio telescopes (ALMA, VLA, ATCA, etc), the bandpass calibration corrects for any effect, therefore the telescope response can be approximated as a gate function centred on the observed frequency and covered by the correlators (which can be discontinuous, depending on the configuration required to the science case)\footnote{two functions in the {\tt mm\_utilities.py} can be found to generate these gate filters}.

\begin{table*} \centering
\caption{Structure of the .fit file. In order, the columns refer to the name of the variable as expressed in the code, 
its type and its purpose.}
\label{tab:setting_file}
\begin{tabular}{l l p{12cm}}
\hline
Parameter & Type & Purposes/Notes \\
\hline \hline
\sourcefile & string &  the source file path where the data are to be read \\
{\it all\_same\_redshift} &  boolean &  keyword to allow for the redshift of the components to be a free parameter \\
\z & float array &  the redshift of each component in the system \\
\modelfile & string & the model file where the parameters of the models and their range are selected (the parameters to fit) \\
\nwalkers & integer & the number of walkers for the fit (should be typical more than 100, emcee documentation advice at least twice the number of free parameters). It play a role on the speed at which the chain converges, so usually the more the better (in the limit of the computer memory). \\
\nsteps & integer & the number of steps for each walkers to perform (can vary widely from 200 to several thousand base on the complexity of the problem) \\
\nstepscut & integer & should be smaller than \nsteps, it represents the number of steps to ignore when drawing the probability density function of the parameters (it depends mainly on convergence time and the number of walkers as it require to select a sufficiently large number of points to populate the parameter space after convergence. \\
\perc & float array & the percentile of the distribution to be returned (used in the plots to show max intervals). Must be odd-dimensioned and all numbers in the range 0$\le$\perc$\le$1. The default values are chosen to represent probability density functions which are not normal, excluding the 10\% outliers. \\
{\it skip\_imaging} & boolean & skip the creation of the data file in images \\
{\it skip\_fit} & boolean & skip the fitting \\
{\it skip\_MCChains} & boolean & skip the generation of the MC-Chains plot \\
{\it skip\_triangle} & boolean & skip the generation of the triangle plot \\
{\it skip\_SED} & boolean & skip the generation of the SED plots \\
{\it unit\_obs} & string & unit of the wavelength/frequency provided in the .dat file \\
{\it unit\_flux} & string & unit of the flux provided in the .dat file \\
\hline
\end{tabular}
\end{table*}

\subsection{Outputs} 
\label{sec:out_fits}

\mrmoose\ produces series of outputs, containing different information (see Table \ref{tab:extension}). The outputs can be divided in two parts: the supporting files (``.fits'', ``.pkl'', ``.sav'', ``.sed'') and the result files (``.pdf''). We focus first on the support files and their contents, allowing to dig further in the interpretation if necessary and allow for an easy reproduction of a given fitting. We designed \mrmoose\ to allow users familiar with the {\tt emcee} package to use their custom tools from the ``.pkl''. However, \mrmoose\ comes also with plotting functions to display results in order to interpret the result from the fit. There are several graphs ``.pdf'' file generated by the procedure: chain convergence, triangle and SED plots. These plots works ``in unison'' to understand the results of the fitting. Any unexpected behaviours in these plots should bring suspicion on the inputs, would it be data, model used, arrangements defined or the settings of the fitting procedure. 

\subsubsection{Supporting files: chains, SED, fitting and fits files}

The ``.fits'' files are generated following the data provided in the ``.dat'' file. For each filter, the procedure generates an image in the FITS format  \citep[commonly used in astronomy][]{Wells1981}, containing all the components associated with this filter, making use of the RA, Dec and resolution values provided in the ``.dat'' file. The procedure creates a unresolved source, assuming a Gaussian point spread function of full width at half maximum provided as the resolution parameter, at the given position (RA, Dec). This allow to check that the combination of the data file are correctly assigned by direct comparison with the original images, if available. We note that the pixel size is fixed to a fifth of the resolution parameter. 

A ``.pkl'' file, contains the object from the {\tt emcee} procedure saved making use of the {\tt cPickle} function from the {\tt pathos} package allowing a serialisation, necessary in case of parallelised use. No modifications are implemented from the original {\tt emcee} sampler, and as such, {\tt emcee}-familiar users can make use of this output into custom graphical tools. This file contains the complete chain for each walker and other diagnostics provided by the {\tt emcee} package.

The ``.sav'' file contains the information from the fitting, including the setting file and the model structures, modified during the fitting and storing the final results in a YAML format. The file contains the final percentiles for each parameter, the best fit value, the name and path of the output files generated by the procedure. 

A ``.sed'' file is created reporting each component of the models from the global best fit. The first column is the frequency in Hz, the following columns reports the flux in erg\,s$^{-1}$\,cm$^{-2}$\,Hz$^{-1}$. 

\subsubsection{Result files: Convergence, Triangle and SED plots}

The convergence plot shows the walkers exploration of the parameter space for each iteration of the procedure for each parameter. This plot helps to identify when convergence is reached, therefore to which value to set the \nstepscut\ parameter to be used by the triangle and SED plots. When is convergence reached? The walkers should oscillate at given values in all windows simultaneously, revealing that they converged into the minimum of the parameter space. Note that several minima can co-exist, therefore the walkers can be split between several values, but as long as the chains are stable around these values (for several hundreds steps), convergence can be considered as reached. However, with the increasing complexity of larger parameter spaces, some walkers can be ``stuck'': i.e. moving erratically, very slowly or not moving at all. Therefore including these walkers in the subsequent plots would bias any interpretation of the probability density distribution. We report for completeness these ``stuck'' walkers in the convergence plot (with a different colour coding) as a check of the fitting (see \S~\ref{sec:examples}). These ``stuck'' walkers will be ignored in the triangle, and SED plots. We filter these out extracting only the chains from the sampler with a high acceptance fraction (meaning that walkers are effectively ``walking''). The default value for this threshold of selection in \mrmoose\ is set to half of the average acceptance fraction value of all chains (AAF value, for instance if AAF = 0.6, the threshold is defined as 0.3 and all chains below are ignored in the following steps). We report both the AAF along with the fraction of walkers filtered out with this cut (SW) in the convergence plot. For specific cases, where the default cut is not optimal, \mrmoose\ allows to set a manual value (see README). As a guide, AAF should be in the 0.2-0.5 range and the stuck walkers fraction relatively low. For better details, we report the reader to the {\tt emcee} documentation. Briefly, in the case of AAF and SW values being extremes, three solutions are to increase the number of walkers (\nwalkers), to better define the initial parameter values through the parameter interval (see \S~\ref{sec:limitations}) or to simply increase the number of steps (\nsteps). 

The triangle plot shows several pieces of information from the fit making use of the {\tt corner} package. It basically uses all information contained by the selected walkers in the chain after the convergence value (\nstepscut) provided in the setting-file. It presents two types of plot: 2D paired-parameter plots and marginalised distribution for each parameter in a form of a triangle. The 2D plots shows the probability density functions by reporting each step of each walker for the projected parameters. Additional contours are also added, representing an alternative version of the dot-representation, with contour values being at the 25\%, 50\% and 75\% of the maximal value of the corresponding marginalised distribution. We stress that these contours do not correspond to the percentiles of the distribution, obtained by cumulative integration of the probability density function. Along the diagonal, the marginalised distributions of each parameter are presented by histograms. The percentile values defined in the setting-file are also reported. We recommend the use of a sightly different version of the seven-figure summary \citep{Bowley1920}, omitting the two extremes, minimum and maximum values of the distribution as particularly sensitive to outliers and ``stuck'' walkers, therefore 10\%, 25\%, 50\%, 75\% and 90\%. They usually present a good summary of different type of distributions (as it is expected to deviate substantially from normal distributions) although should not prevent to carefully check each distribution visually to identify problems. This triangle diagram is particularly helpful to (i) reveal any degeneracies among parameters (ii) reveal if parameters are constrained, and how well, given the dataset. 

The SED plots show the fit of each arrangement on the data, the final results of the models and data combination. The final number of generated plot changes, depending on the complexity of the source considered. First, the SED is coming in two versions, a best fit and a ``spaghetti'' version. The former shows the best overall fit to the data, each colour corresponding to one model (the maximum likelihood reached in the chains). The latter shows all the solutions provided by each walkers at each step after the convergence cut (same than the triangle plot). This is particularly useful to see how good the models are constrained with the data in combination with the triangle and convergence plots. We choose this representation over a shaded area representation to better visualise the constraints from the data and to better highlight the different solution in case of the presence of several minima. For sources with several arrangements, split versions of the SED will also be produced to ease interpretation (see examples in \S~\ref{sec:examples}) with colour coding of the models conserved over the windows.

\begin{table*} \centering
\caption{List of the output files and their use.}
\begin{tabular}{l p{14cm}}
\hline
Name & Purposes/Notes \\
\hline \hline
triangle\_plot & probability density functions for each parameter and by pair. This plot is particularly efficient to reveal degeneracies between parameters (banana shape) 
or unconstrained parameters by the fit (no single peak in the probability density function) \\
MCChains\_plot & values of parameters at each step for each walker. This plot is particularly useful to show when convergence is reached, used to calculate all the quantities displayed in other plots \\
SED\_fnu\_plot & plot in units of frequency and flux of all the data and models on the same graphical window. While relevant for few components or arrangements, this plot becomes rapidly too complex for multi-component case \\
SED\_fnu\_splitplot & split plot in unit of frequency and flux for each arrangement defined in the data file \\
SED\_fnu\_spaghettitplot & plot showing all the possible parameter values on the SED after convergence (at each step for each walker). This plot translates the triangle plot on the data, showing how constrained the model is compared to the data. \\
SED\_fnu\_splitspaghettitplot & equivalent to the SED\_fnu\_splitplot but for the spaghetti style \\
SED\_file & file storing the SED of each component as plotted in the SED\_fnu plot \\
sampler\_file & file storing the chains of the fitting.  \\
save\_struct & file storing the details of the fit for future references and reproducibility, save the fit settings dictionnary \\
\hline
\end{tabular}
\end{table*}

\section{Examples}
\label{sec:examples}

In order to demonstrate \mrmoose\ capabilities, we focus here on three specific extragalactic examples which where used to build this procedure (the code is provided with several examples with increasing complexity, we focus here on the artificial ones produced by the file {\tt example\_1.py} and {\tt example\_6.py}) as well as a previous result from literature. For the sake of simplicity, we will refer to them as example \#1, \#2 and \#3, respectively. The first is a simple case of fitting, with only one model onto a dataset, here synchrotron emission from a radio source in the 74\,MHz-8.4\,GHz range. The second case is much more complex: we here aim to disentangle the contribution of various components, spatially non-collocated and originating from different physical processes. We take the example of a powerful radio galaxy and a star forming galaxy companion with separate emission from the lobes (of synchrotron origin), the dust heated by star formation and the dust heated by AGN, in the 8\mum-1.4\,GHz wavelength range. The third example cover a real dataset published in \cite{Gilli2014}, fitting a modified black body in the 70\mum-3\,mm range to reproduce dust emission of a distant quasar, demonstrating the useful addition of upper-limits in the fitting process.

\subsection{Example \#1: single source, single component}
\label{sec:ex1}

This example is the simplest fitting imaginable for \mrmoose, consisting of a single law on a given dataset. Although of limited interest, it allows us to familiarise with the code. We benchmark \mrmoose\ by generating a fake source and trying to recover these original parameters. 
 
\subsubsection{Fake source and input files}

To generate a fake source data-points, we assume a single, unresolved source at $z=0$ emitting a single synchrotron law represented by the model {\tt sync\_law} with $\alpha$=-1 and log$_{10}N$=-22 (to recover a source with flux at Jy-level) where $\alpha$ is the slope and $N$ the normalisation (in log-scale). We generate five single points passing this SED through the filters from the filter library and assign a given signal-to-noise ratio (here {\it SNR}=5 for all points). We simulate further observation uncertainties by adding a random gaussian noise with a standard deviation corresponding to the provided SNR value. In practice, the code adds (or subtracts) flux to each datapoint, not following exactly the original synchrotron law defined previously. By adding this extra noise the chance of recovering the exact parameter values decreases, this perfectly illustrates the real case of observations, where each point is drawn from a distribution. This fake data-file (Listing~\ref{lst:data1}) is, along with the setting-file (Listing~\ref{lst:fit1}) and the model-file (Listing~\ref{lst:mod1}), generated from the {\tt example\_1.py} procedure (provided with \mrmoose). These files are therefore fed into \mrmoose\ to perform the SED fitting. 

\subsubsection{Output files and results}

\begin{figure}
\includegraphics[width=0.5\textwidth]{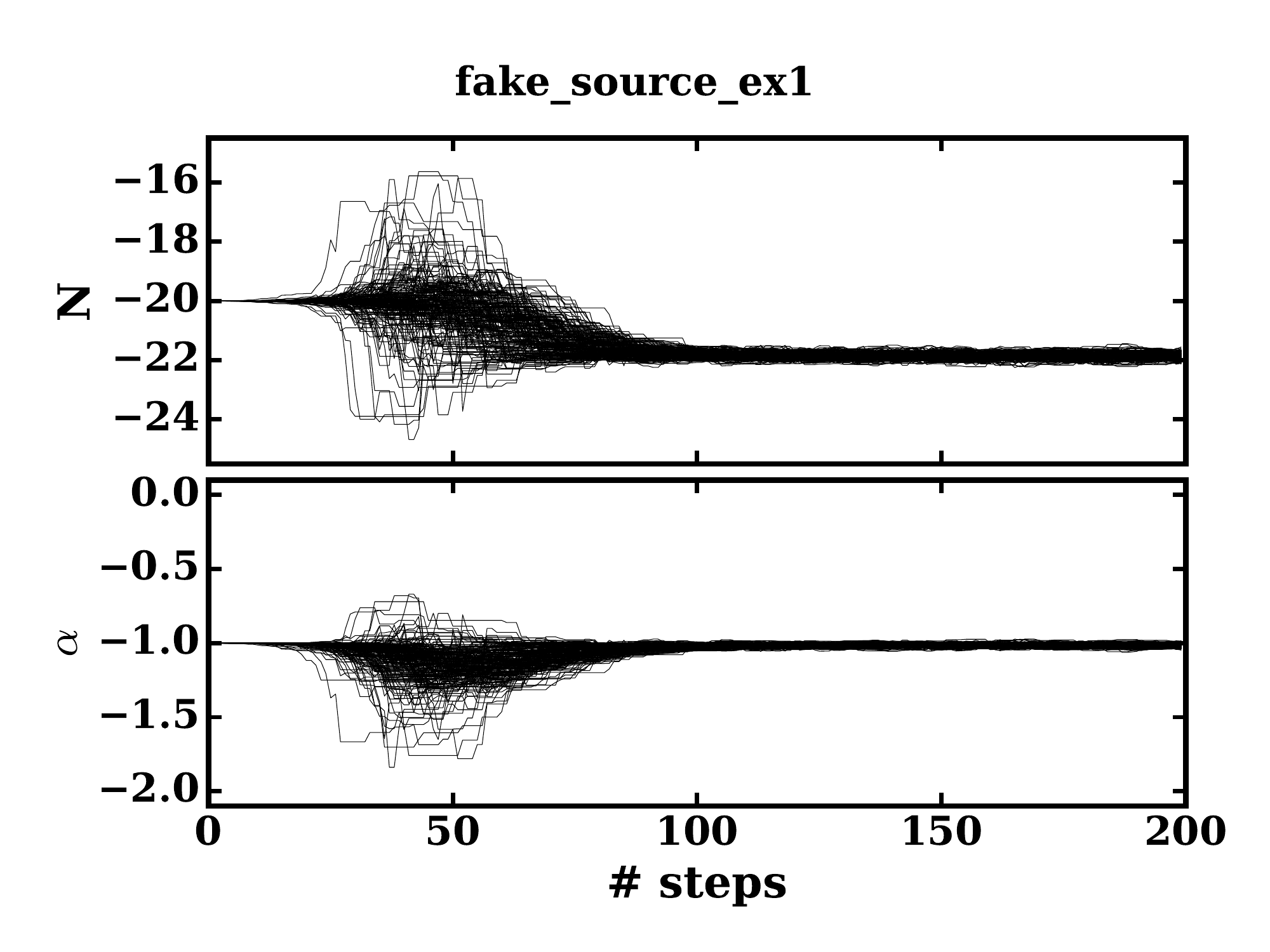}
\caption{MC-Chains for each parameter for the given number of steps for example \#1. Note the convergence reached roughly after step \#100}
\label{fig:ex1_mcchains}
\end{figure}

\begin{figure}
\includegraphics[width=0.5\textwidth]{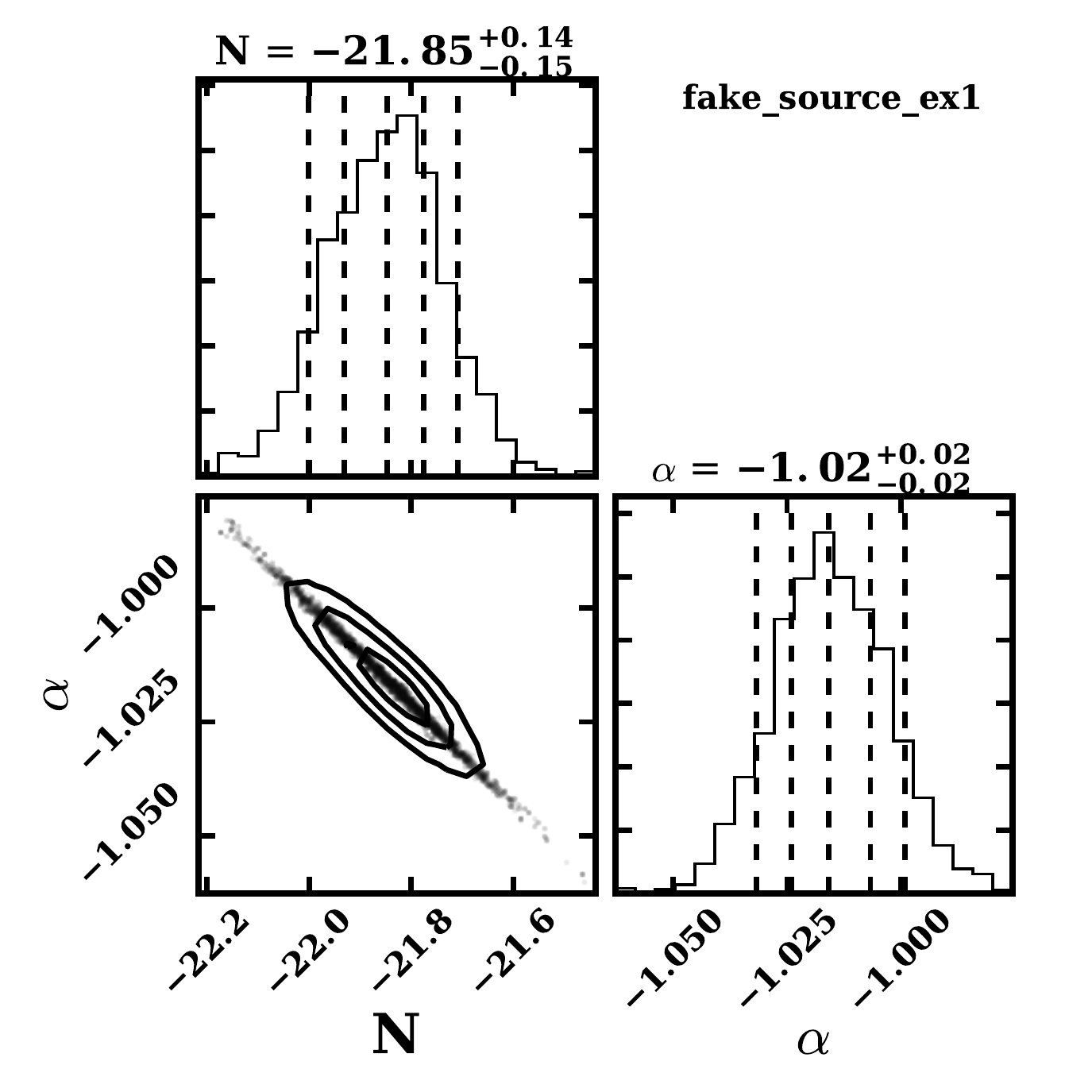}
\caption{Triangle diagram for example \#1. The diagonal represents the marginalised probability distribution. The vertical dashed line correspond to the 10\%, 25\%, 50\%, 75\% and 90\% percentiles of the distribution. The 50\% percentile parameter along the 25\% and 75\% percentile values are reported in Table \ref{tab:ex1_param}. The other diagram represents the 2D projection of the parameter space on the corresponding parameters on the abscissa and ordinate axes. The contours represents the density distribution at 75\%, 50\% and 25\% of the maximum parameter value . }
\label{fig:ex1_triangle}
\end{figure}

\begin{figure*}
\centering
\begin{tabular}{cc}
\includegraphics[width=0.48\textwidth]{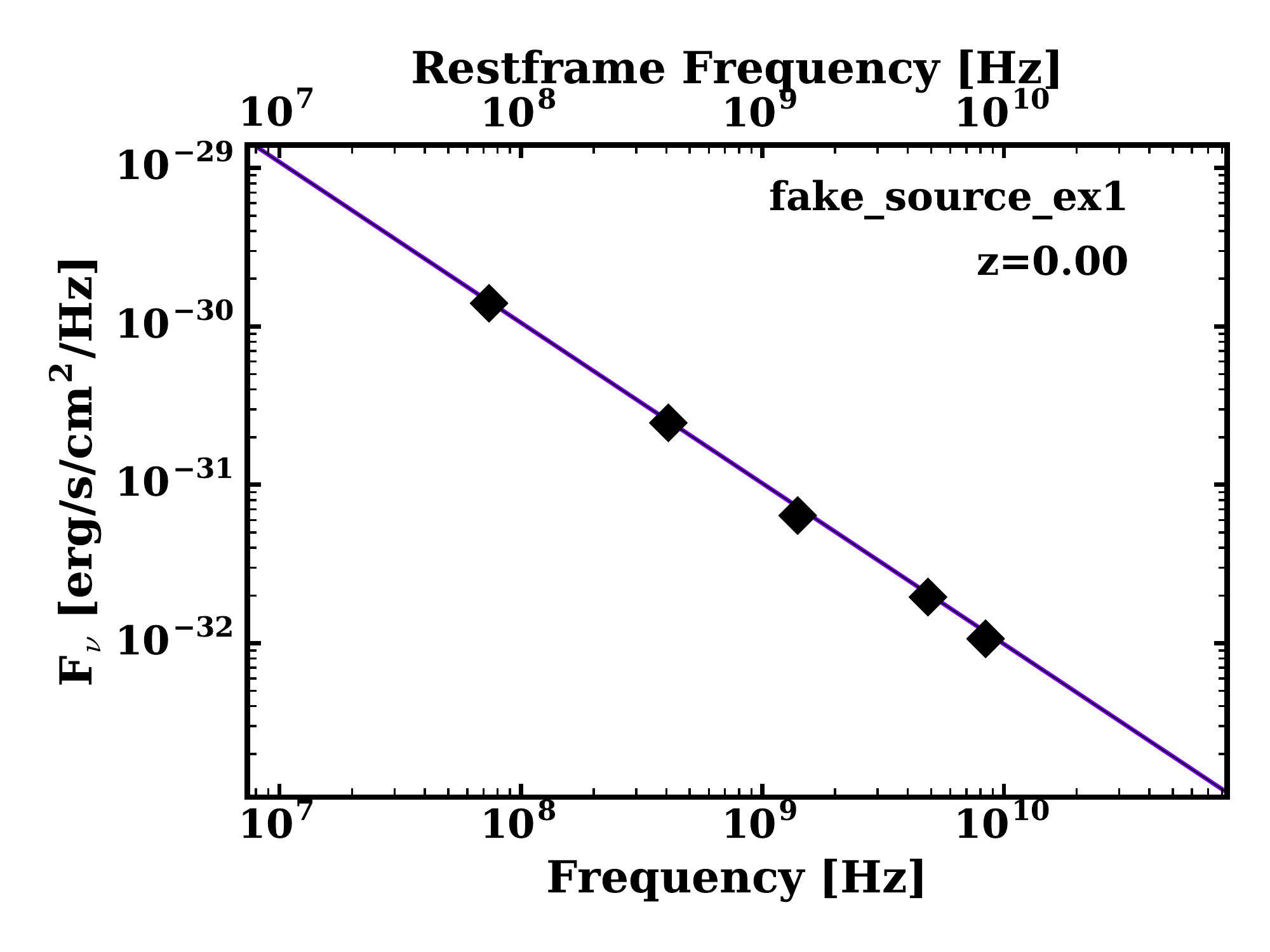} &
\includegraphics[width=0.48\textwidth]{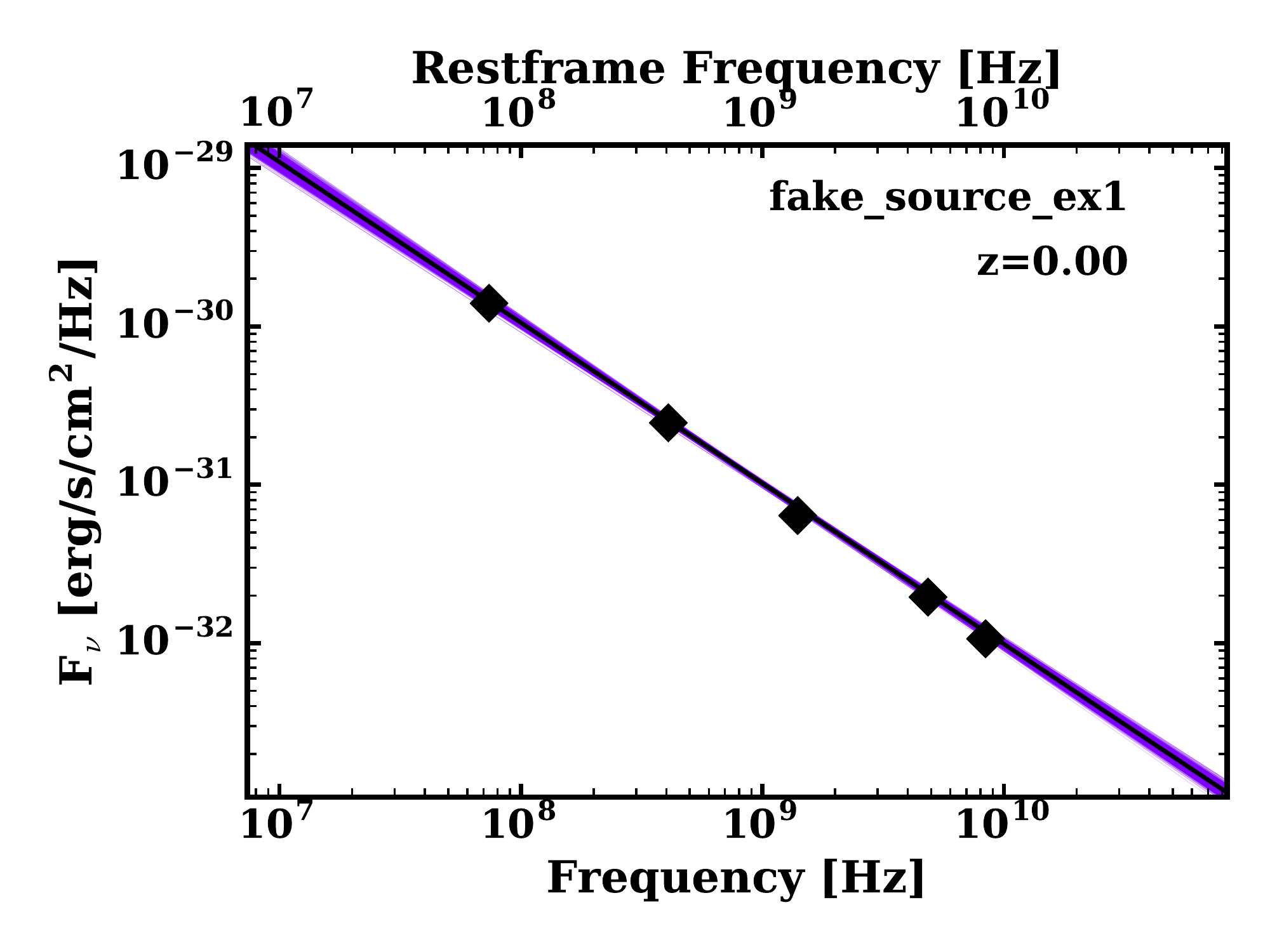} \\
\end{tabular}
\caption{SEDs for example \#1. Note that the uncertainties are plotted but are smaller than the datapoints,  diamonds being detections. {\it left:} best fit plot. The plain black line is the sum of the different components in each arrangement. {\it right} ``spaghetti'' plot. The best fit here is reported as the black plain line for each component (defined in the .dat file). The shaded area is the sum of each model representation after convergence (values of the parameters at each step of each walker).}
\label{fig:ex1_sed}
\end{figure*}

\begin{table} \centering
\caption{Components and parameters of example \#1. $^*$ value in logarithmic scale. The uncertainties quoted are the 25 and 75 percentiles.}
\label{tab:ex1_param}
\begin{tabular}{cccc}
\hline
Name & parameter & true value & fit value \\ 
\hline \hline 
\multirow{2}{*}{sync\_law} & $N$ & -22$^*$ & -21.85$^{+0.14}_{-0.15}$\\ [0.2ex]
                                  & $\alpha$ & -1 & -1.02$^{+0.02}_{-0.02}$\\
\hline
\end{tabular}
\end{table}

The code generates the output files previously described (see \S~\ref{sec:design}). We focus on the four ``.pdf'' plots: the convergence plot (Fig. \ref{fig:ex1_mcchains}), the triangle plot (Fig. \ref{fig:ex1_triangle}) and two SED plots (in unit of $F_{\nu}$, Fig. \ref{fig:ex1_sed}). 

\noindent{The} MC-Chains plot (Fig. \ref{fig:ex1_mcchains}) shows here a quick convergence (reached roughly at the \#100 step). We define \nstepscut=180, and apply it to the subsequent plots (triangle and SEDs). 

The triangle plot (Fig. \ref{fig:ex1_triangle}) is the classical plot to represent multi-parameter space: plotting the distribution of parameters in pairs with the marginalised distribution of each parameter on the hypotenuse of the the triangle. Along this diagonal, the marginalised distribution for each parameter is a histogram. The dashed lines are the percentiles provided in the setting-file, here 10\%, 25\%, 50\%, 75\% and 90\% (provided by the user in the setting-file), as recommended in \S~\ref{sec:design}. The parameters are well constrained in a small area of the parameter space. It is interesting to note that even if the two marginalised distributions show relatively Gaussian-like distributions, the 2D projection is far from being a symmetric cloud. The contours correspond here to 25\%, 50\% and 75\% from the maximal value of the marginal distributions. Overall, we note that the input parameter are recovered within the uncertainty interval (remembering we added extra noise to the data, see Table~\ref{tab:ex1_param}).

As presented in \S~\ref{sec:design}, the code generates two version of the SED to visualise convergence and the effect of uncertainties: the best fit and the ``spaghetti'' version allows to visualise the how well the data constrain the parameters of the model. Fig. \ref{fig:ex1_sed} shows that the power-law parameters are well constrained with only a small scatter of the model in the data-points.
 
\subsection{Example \#2: multi-sources, multi-components}
\label{sec:ex2}

\begin{figure}
\includegraphics[width=0.5\textwidth]{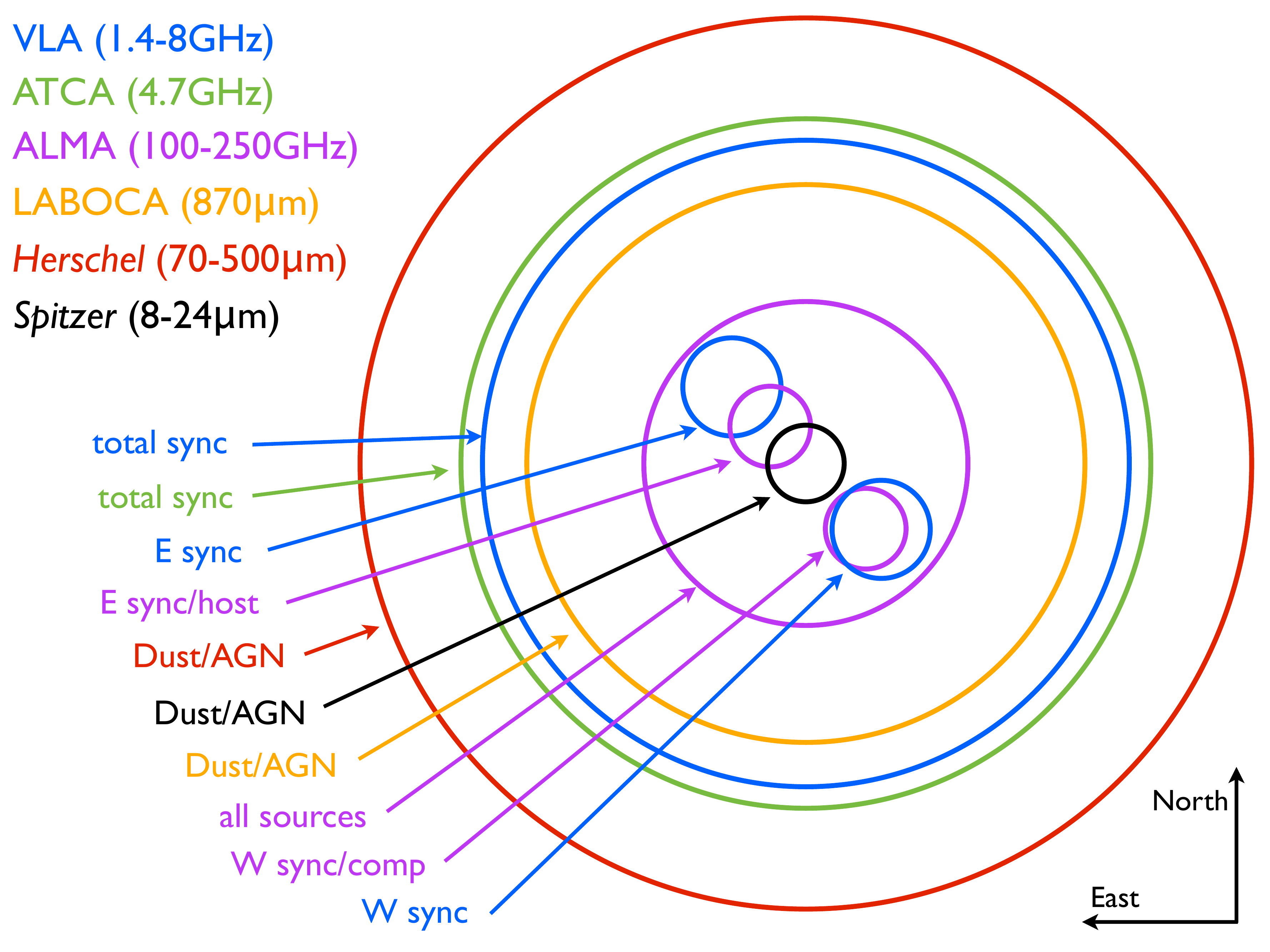}
\caption{Scheme of the combination of observations at different wavelengths from the data file to illustrate the considered example. 
The circle sizes are relative to the resolution of the observation but not to scale (see Fig.~\ref{fig:stamps}). Note that {\it Herschel}, LABOCA, one ALMA, one VLA and ATCA observations are not resolving the different components. From bigger to small circles: \herschel, ATCA, VLA, LABOCA, ALMA, double circles are VLA and ALMA(closer to center), and in the center \spitzer.}
\label{fig:spatial_sed}
\end{figure}

For the second example, we aim to demonstrate the application of \mrmoose\ to a much more realistic situation where the total emission received is a complex combination of several blended sources at different wavelengths by using all the variations and freedom allowed by \mrmoose. This fake system is adapted from a real science case, see \citet{Gullberg2016}. In the following paragraph and Fig.~\ref{fig:spatial_sed}, we summarise this complex source consisting of two galaxies, one powerful radio galaxy with two bright radio lobes \cite[FRII type, ][]{Fanaroff1974} and a star-forming companion galaxy. 

\begin{table} \centering
\caption{Components and parameters of example \#2. $^*$All normalisation values are logarithmic values. The quoted uncertainties are the 25 and 75 percentiles.}
\label{tab:ex2_param}
\begin{tabular}{cccc}
\hline
Name & parameter & true value & fit value \\ 
\hline \hline 
\multirow{2}{*}{AGN\_law} & $N_{\rm AGN}$ & -32$^*$ & -31.94$^{+0.08}_{-0.09}$ \\
                                  & $\alpha_{\rm AGN}$ & -2 & -2.04$^{+0.13}_{-0.12}$ \\ [1ex]
\multirow{2}{*}{BB\_law(1)} & $N_{\rm BB1}$ & -22$^*$ & -21.75$^{+0.29}_{-0.22}$ \\
                                 & $T_{1}$ & 40 & 42.98$^{+13.30}_{-14.82}$ \\ [1ex]
\multirow{2}{*}{BB\_law(2)} & $N_{\rm BB2}$ & -22$^*$ & -21.74$^{+0.30}_{-0.18}$ \\
                                 & $T_{2}$ & 60 & 45.51$^{+9.71}_{-16.63}$ \\ [1ex]
\multirow{2}{*}{sync\_law(1)} & $N_{\rm s1}$ & -22$^*$ & -21.64$^{+0.93}_{-0.87}$ \\
                                    & $\alpha_{\rm s1}$ & -1 & -1.06$^{+0.09}_{-0.10}$ \\ [1ex]
\multirow{2}{*}{sync\_law(2)} & $N_{\rm s2}$ & -22$^*$ & 21.88$^{+0.91}_{-0.87}$ \\
                                    & $\alpha_{\rm s2}$ & -1.2 & -1.03$^{+0.09}_{-0.10}$ \\ [1ex]
\hline
\end{tabular}
\end{table}

\subsubsection{Fake source and input files}

\begin{figure*}
\includegraphics[width=0.8\textwidth, trim= 0 50 0 30]{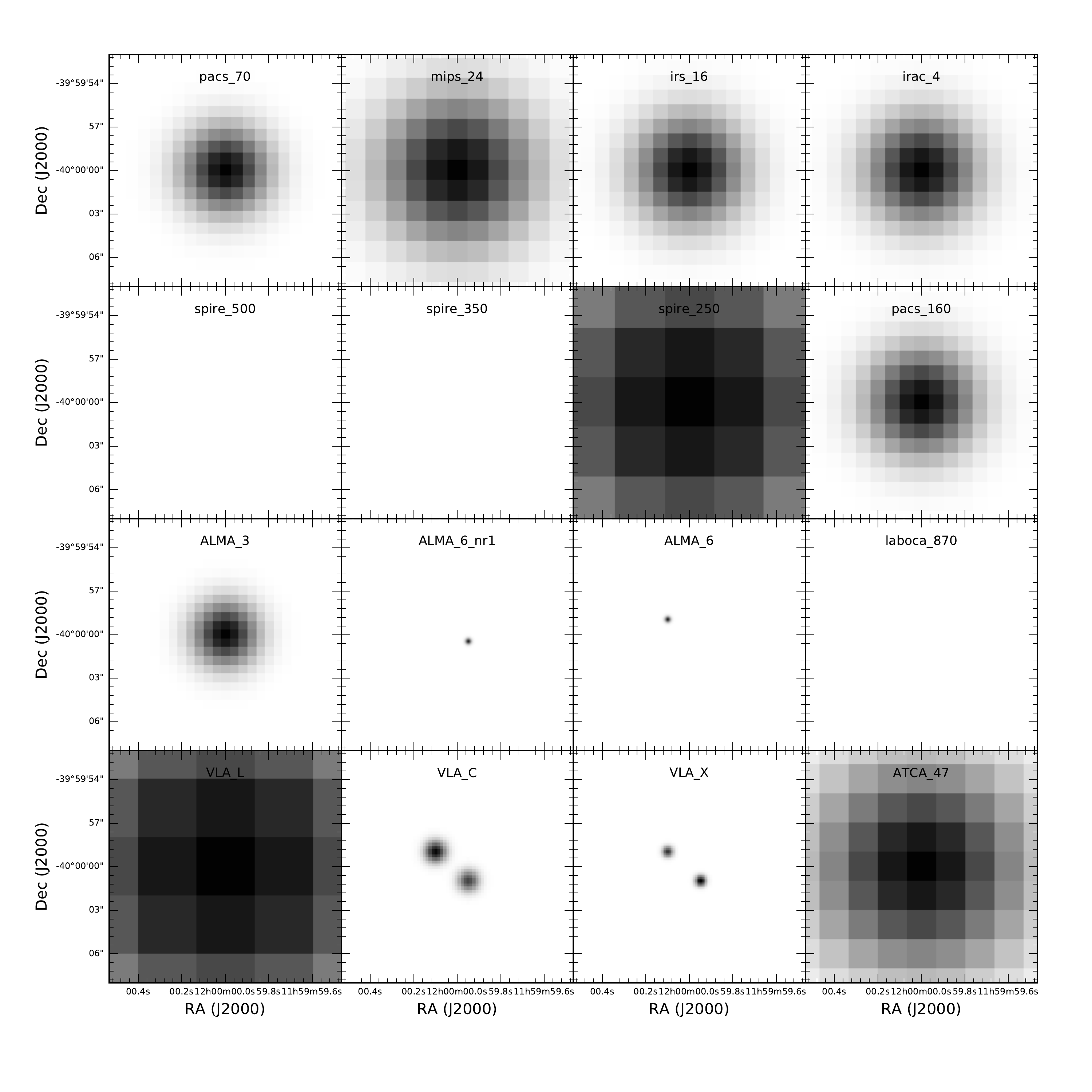}
\caption{Images generated from the information contained data file assuming a Gaussian point spread function, all images scaled to the same coordinates. Frequency is decreasing toward the right and the top. Note that no sources are detected in {\it spire\_350, spire\_500 and laboca\_870} images, while {VLA\_X and VLA\_C} present two components. North is top and East is left on all images.}
\label{fig:stamps}
\end{figure*}

The following characteristics and assumptions are used to define this system: 
\begin{itemize}
\item The radio galaxy is composed of an AGN, a host galaxy and is star-forming
\item The AGN is radio-loud presenting a FRII morphology (two bright radio lobes)
\item The companion is star-forming and without AGN
\item The SED coverage covers the 8\mum-1.4\,GHz range with various resolutions and detections as the result of the different facilities used for observations.
\end{itemize} 
Fig. \ref{fig:stamps} presents the observations, each image generated by \mrmoose\ before the fitting. To further precise, the IR SED do not differentiate the two galaxies due to the lack of resolution and is constituted of three components: the cold dust from both the companion and the host and the AGN hosted in the radio galaxy. We also note that the 350\mum, 500\mum\ and 870\mum\ are only upper limits and therefore, the sum of these three components cannot largely exceed these values. The synchrotron emission is constituted of two lobes, partially blended with the radio galaxy itself and with the companion or unresolved at the lowest radio frequency. Also, the synchrotron and dust emission are equally contributing at certain frequencies (here 230\,GHz). 

We have a total of 14 observations in the 8\mum-1.4\,GHz range, with 17 photometric data-points. The total number of components is therefore five, one AGN dust component, two cold dust component (one the radio galaxy, one of the companion galaxy assuming that the cold dust component originates from young stars) and two synchrotron emissions (originating from each lobe). This corresponds to ten free parameters, summarised in Table~\ref{tab:ex2_param}. Given the combination of resolution, sensitivity, frequency coverage, photometry and components, we can define seven arrangements of the data/models, used to generate the fake data-file and model-file.
\begin{enumerate}
\item Dust/AGN: combination of the dust emission from the companion and the radio galaxy, itself consisting of the AGN and cold dust components {\it Spitzer}, {\it Herschel} and LABOCA \citep{Siringo2009}:
\item W sync/comp (western synchrotron, companion): partially resolved at 229\,GHz, combination of synchrotron and cold dust emission (ALMA Band 6, 7.5\,GHz bandwidth)
\item E sync/host (eastern synchrotron, host galaxy): partially resolved at 245\,GHz, combination of synchrotron and cold dust emission (ALMA Band 6, 7.5\,GHz bandwidth)
\item total sync (total synchrotron): unresolved radio emission at 1.4\,GHz, 4.7\,GHz 102\,GHz (VLA, ATCA and ALMA Band 3 at 50\,MHz, 2\,GHz and 7.5\,GHz bandwidth, respectively)
\item W sync: only resolved at 4.8\,GHz and 8.4\,GHz (VLA, C and X bands, with 50\,MHz bandwidth)
\item E sync: only resolved at 4.8\,GHz and 8.4\,GHz (VLA, C and X bands, with 50\,MHz bandwidth)
\item total emission: combination of all components at 102\,GHz (ALMA Band 3, with 7.5\,GHz bandwidth)
\end{enumerate}

We generate the fake files using the fake source procedure provided with \mrmoose, generating, summing and passing the different combination through the adequate filters, in the same fashion presented in \S~\ref{sec:ex1}. Table~\ref{tab:ex2_param} lists the values of each parameter and Fig. \ref{fig:stamps} the images corresponding to each filter from the data-file. We choose \nwalkers=400 to make sure to have a sufficiently large number of walkers exploring the parameter space and a deliberately (too) large number of steps (\nsteps=3500) to ensure convergence is reached. 

\subsubsection{Output files}
\label{sec:ex2_out}

\begin{figure*}
\includegraphics[width=\textwidth,trim= 0 30 0 10]{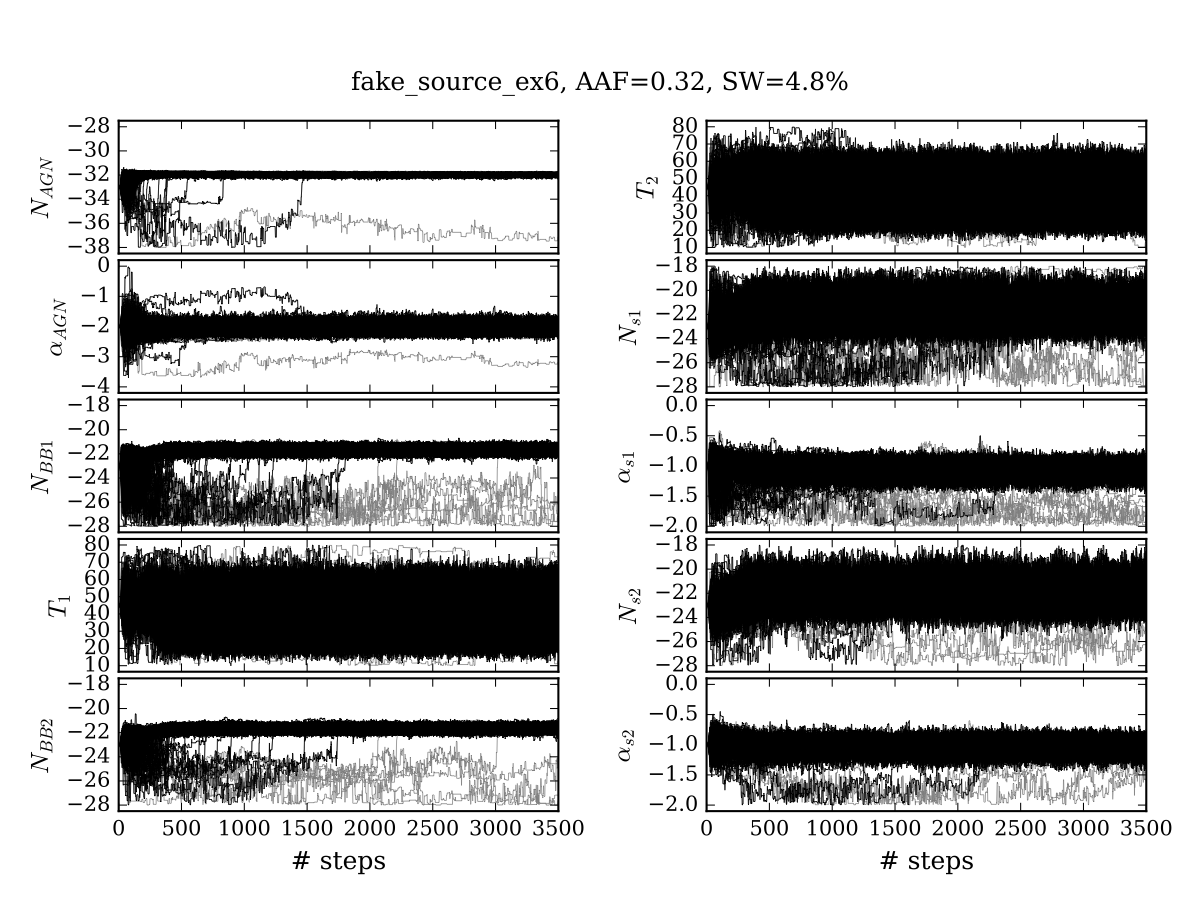}
\caption{MC-Chains for each parameter for the given number of steps for example 2. Note the convergence reached roughly after 1000 steps. The grey lines are the ``stuck'' walkers, excluded from further analysis (see \S~\ref{sec:ex2_out} and Fig. \ref{fig:histo}).}
\label{fig:ex2_mcchains}
\end{figure*}

The code generates the output files previously described in \S~\ref{sec:design}. Given the higher number of arrangements, two new SED files are generated (compared to example 1), presenting a split version of the total SED. Indeed, when dealing with complex system, presenting emission from different origins, over-plotting everything on the same SED makes interpretation difficult. Therefore, \mrmoose\ provides split SEDs, with the same scale and range for each arrangement (seven in this example) with color coding each model component. We describe here the results from the convergence plot (Fig. \ref{fig:ex2_mcchains}), the triangle plot (Fig. \ref{fig:ex2_triangle}) and finally finish by the SEDs (Fig.~\ref{fig:ex2_singlesed}-\ref{fig:ex2_splitsed}).

\begin{figure}
\includegraphics[width=0.5\textwidth, trim= 0 30 0 30]{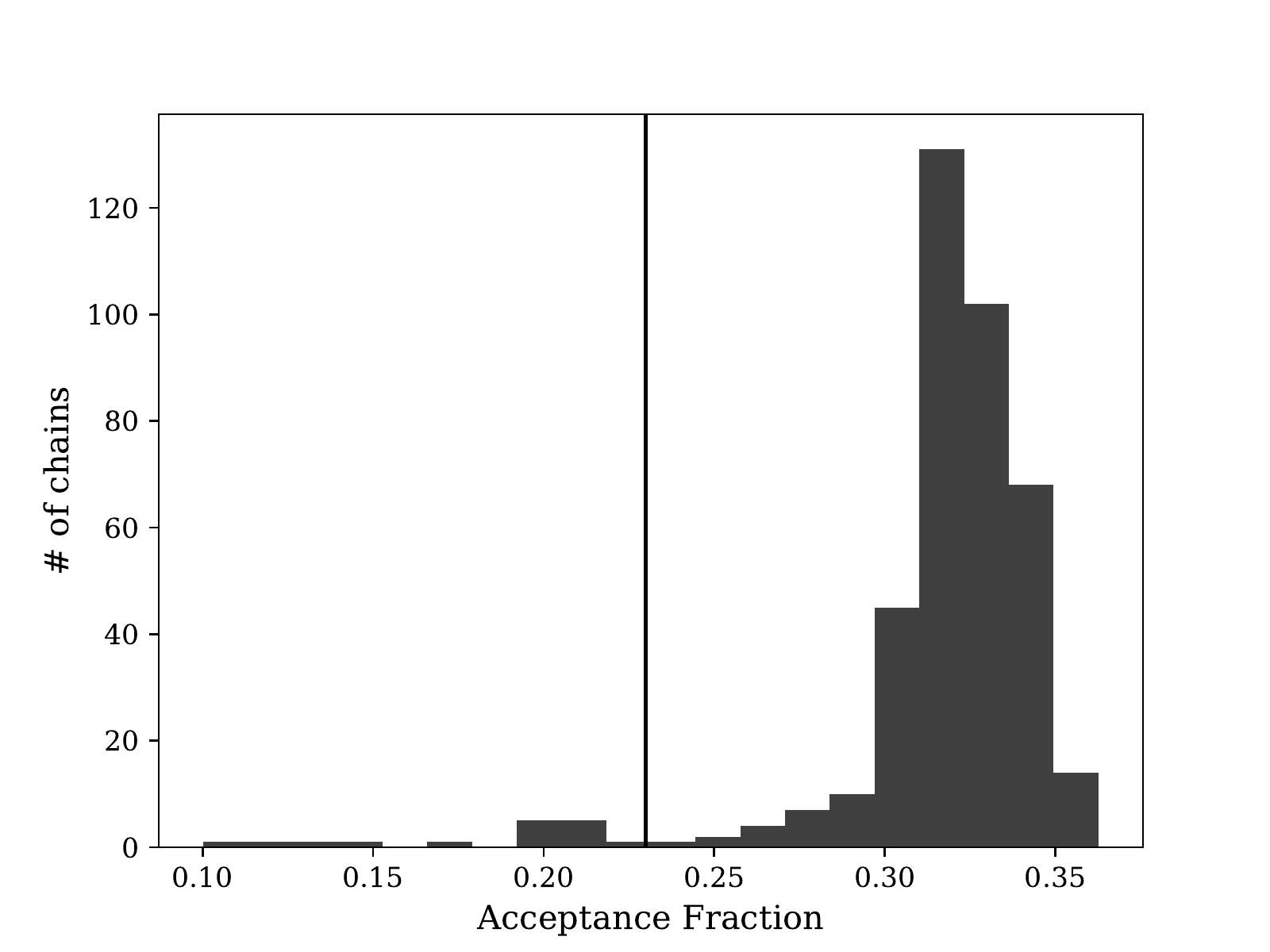}
\caption{Histogram of the acceptance fraction of the 400 chains. Note the cut at 0.23, chains under this values are reported in grey in Fig. \ref{fig:ex2_mcchains}, which are the ``stuck walkers'' (see text \S~\ref{sec:ex2_out}).}
\label{fig:histo}
\end{figure}

\begin{figure*}
\includegraphics[width=\textwidth,trim= 100 45 120 35]{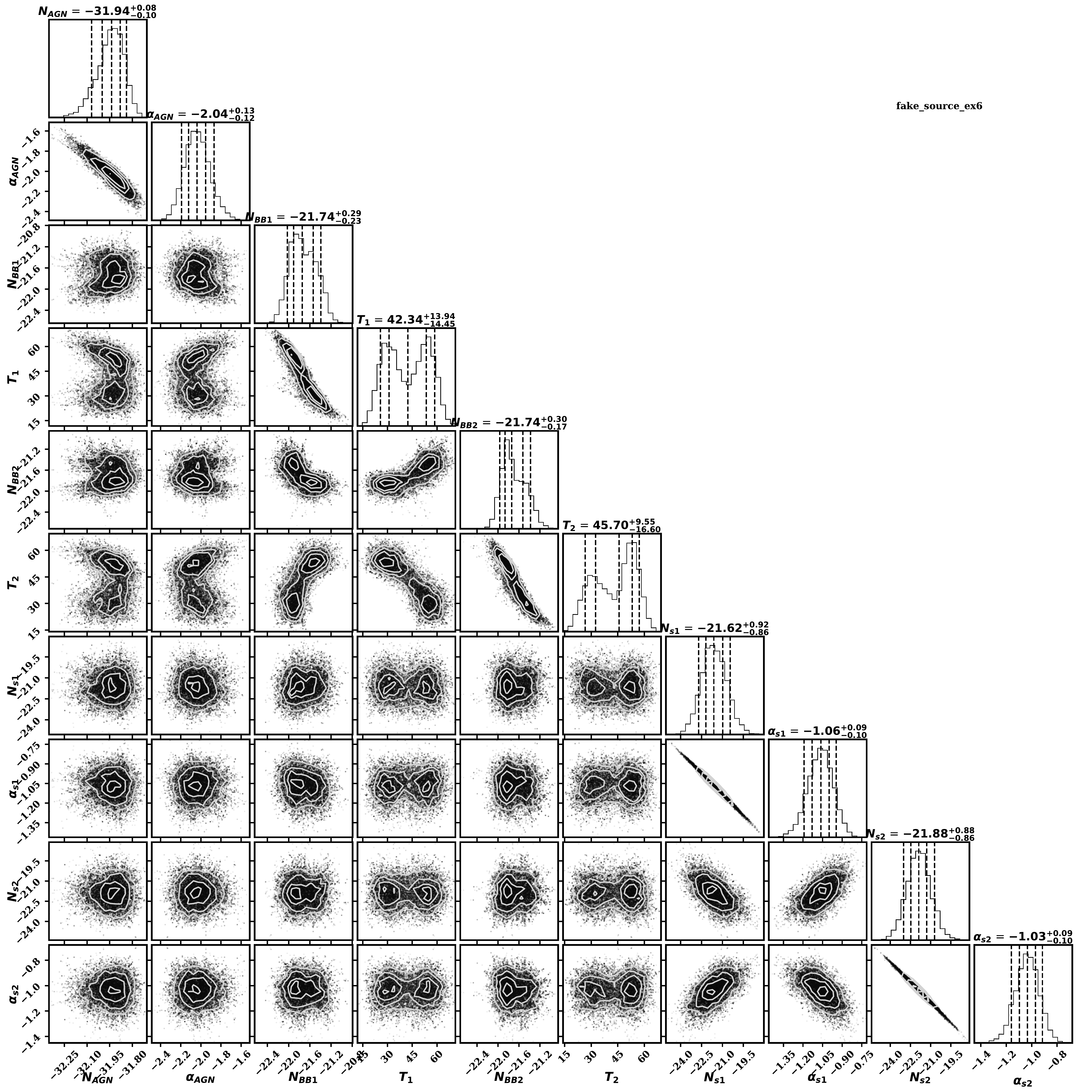}
\caption{Triangle diagram for example \#2. The diagonal represents the marginalised probability distribution. The vertical dashed line correspond to the 10\%, 25\%, 50\%, 75\% and 90\% percentiles of the distribution. The 50\% percentile parameter along the 25\% and 75\% percentile values are reported in Table \ref{tab:ex2_param}.The other diagrams represent the 2D projection of the parameter space on the corresponding parameters on the abscissa and ordinate axes. The contours represents the 75\%, 50\% and 25\% of the maximum parameter value. }
\label{fig:ex2_triangle}
\end{figure*}

\begin{figure} \centering
\includegraphics[width=0.5\textwidth,trim= 0 40 0 20]{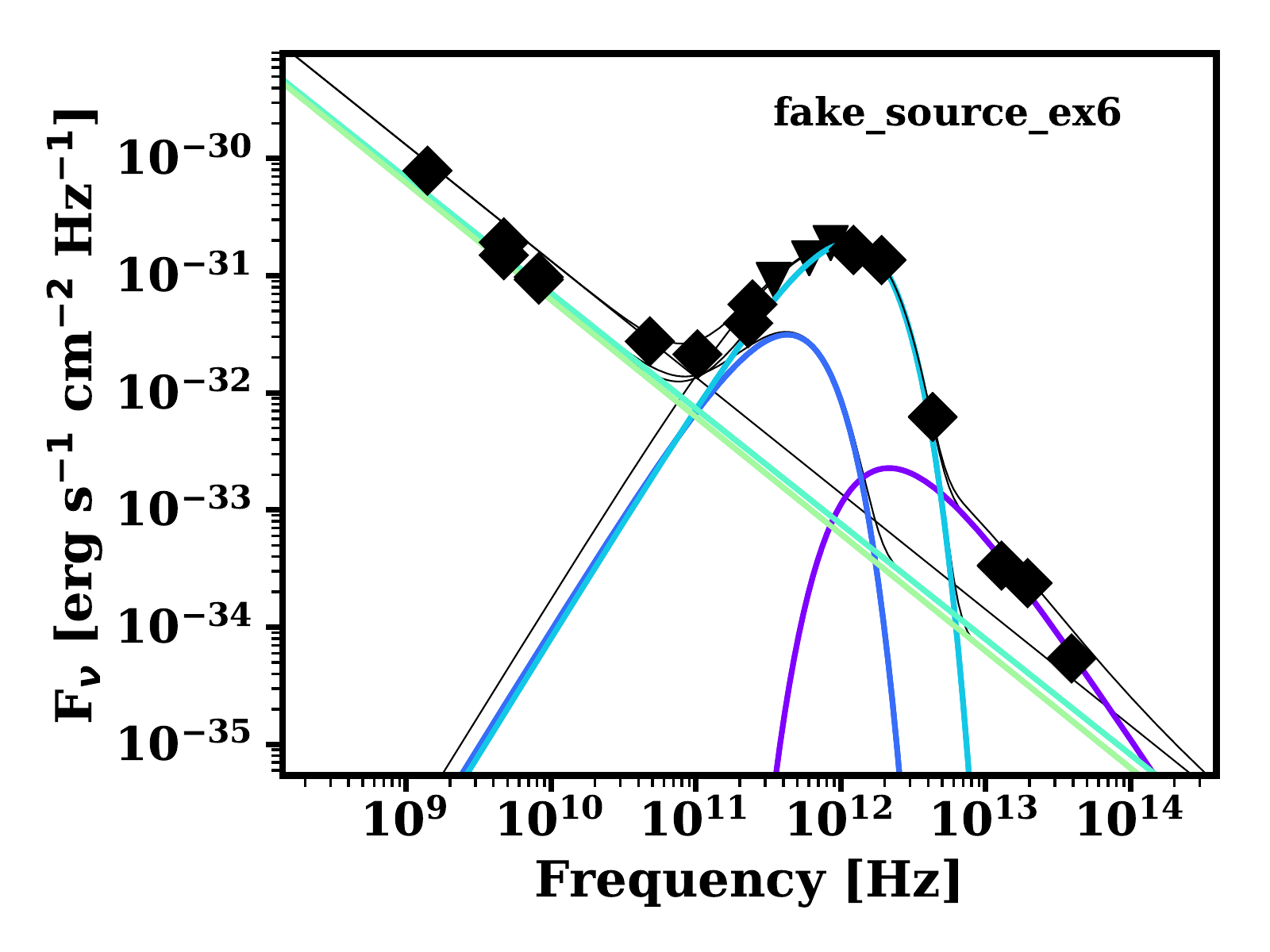}
\caption{Total SED for each arrangement of data and models plot in a single figure. Coloured lines refer to the different models and the black lines are the total of all components in each arrangement. This figure is reported for pedagogic purposes only as, given the increasing complexity of the system, a split SED is much clearer to disentangle the various contribution of the components to the data (see Fig.~\ref{fig:ex2_splitsed}).}
\label{fig:ex2_singlesed}
\end{figure}

\begin{figure*} \centering
\begin{tabular}{c}
\includegraphics[width=0.85\textwidth]{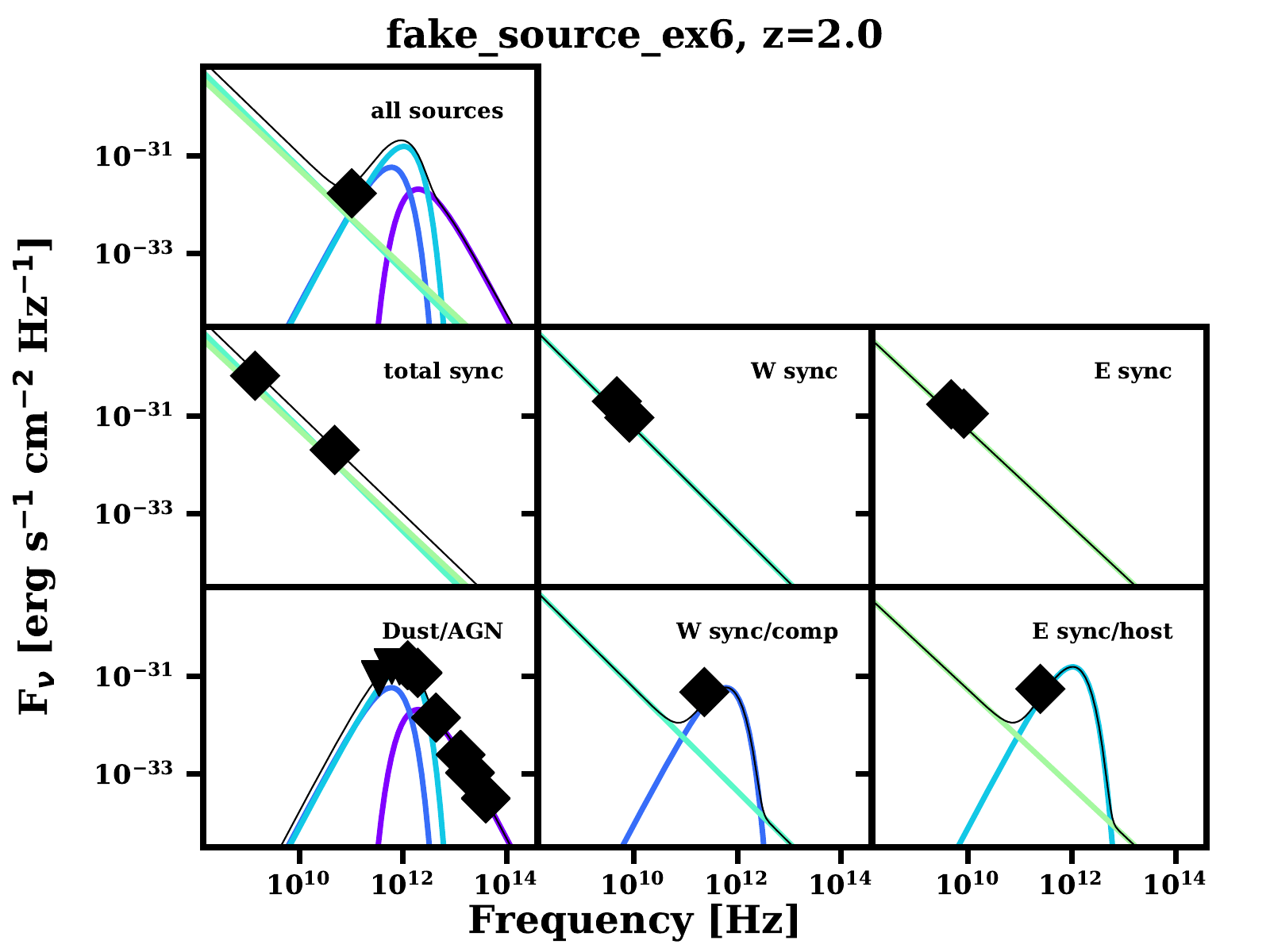} \\
\includegraphics[width=0.85\textwidth]{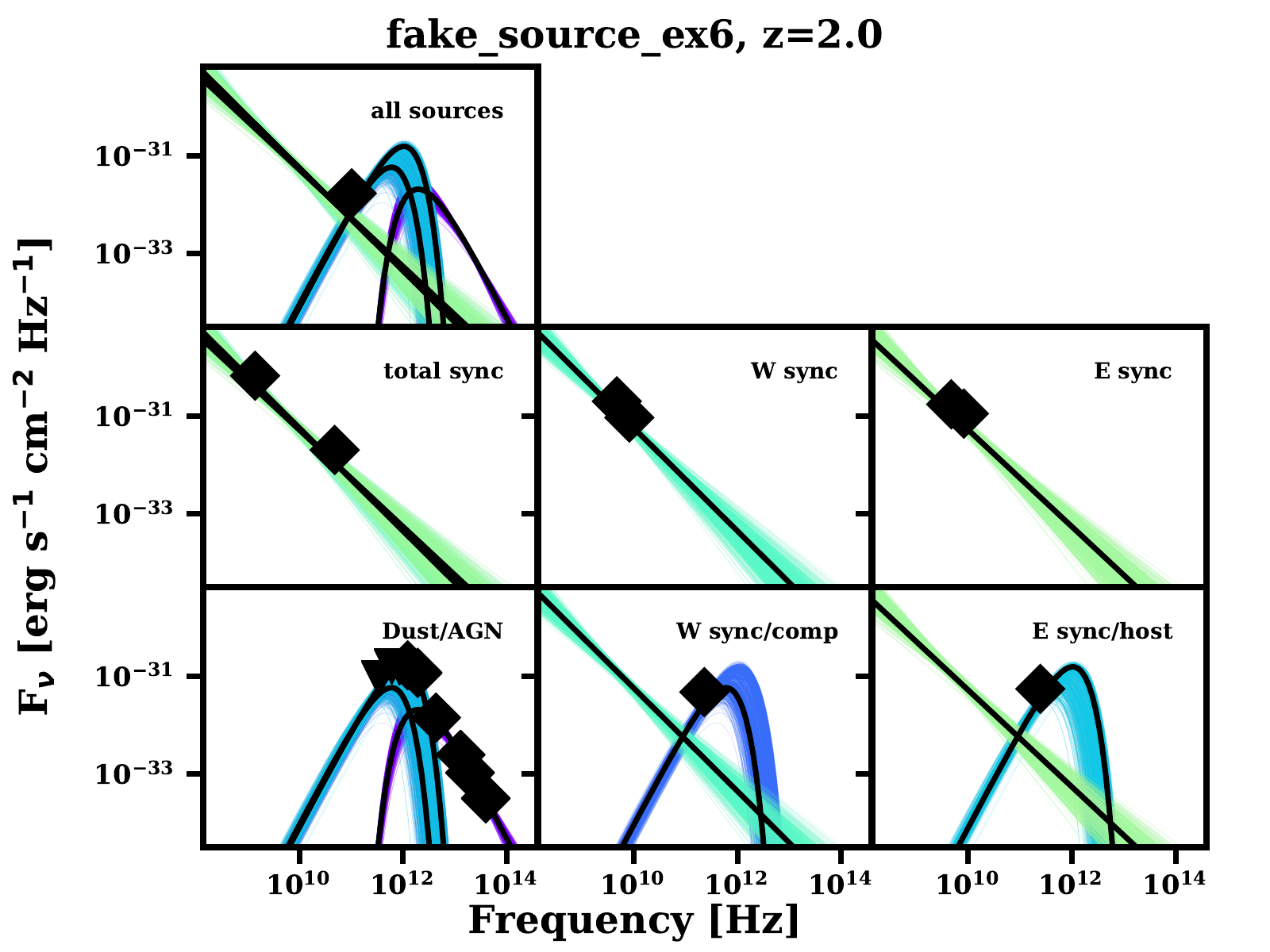} \\
\end{tabular}
\caption{SEDs for example \#2, split into subplots to illustrate each arrangement. Note that the uncertainties are plotted but are smaller than the datapoints. Diamonds are detections while downward triangle are upper limits. {\it top:} best fit plot. The plain black line is the sum of the different components in each arrangement. {\it bottom} ``spaghetti'' plot. The best fit here is reported as the black plain line for each component (defined in the .dat file). The colours refer to each model component, consistently between each subplot.}
\label{fig:ex2_splitsed}
\end{figure*}

Fig. \ref{fig:ex2_mcchains} shows the chains of walkers, where convergence is reached roughly after 1000 steps. With more attention, one notices each parameter converges at different times, the first ones being the most constrained parameter such as $N_{\rm AGN}$ or $\alpha_{\rm AGN}$ (compared to $N_{\rm BB1}$). Also from this plot, it is clear that the dust temperature cannot be further constrained without extra information. We deliberately chose a larger number of steps to ensure we would not miss a long convergence trend. The grey chains corresponds to flagged walkers, below the acceptance fraction threshold discussed in \S~\ref{sec:design}. In this particular example, the default AAF threshold value does not filter out some isolated walkers which did not converge even after this large number of steps. This is probably due to the increasing complexity of the parameter space, where several local minima coexist. We therefore set a manual acceptance fraction threshold at 0.23. We choose this value because it filters out the low-end tail of the distribution of the acceptance fraction values (see Fig. \ref{fig:histo}) corresponding to the isolated walkers in the parameter space (see the bottom of the $N_{\rm BB2}$ chains, at roughly step \#1000). To plot the following SEDs and triangle diagram, we set the limit \nstepscut=3450. A much lower cut, such as 3000, is possible but would clutter the following diagrams and given our 400 walkers, this still represents 20000 points to sample our distributions.

Fig. \ref{fig:ex2_triangle} provides us with information on the parameters. Looking at the histogram on the diagonal, we can see that most parameters are relatively well constrained, within the original values, except for the temperature of the cold dust components, especially $T_1$ (This will be particularly obvious in the SED plot). We also see that the increasing complexity of the system, contours are very useful to highlight ``banana'' shapes more clearly. This is a perfect demonstration of how \mrmoose\ provides reliable results, in the ambiguous case of a lack of data. 

Fig. \ref{fig:ex2_singlesed} and Fig. \ref{fig:ex2_splitsed} respectively show the single and the split SEDs for the best fit and ``spaghetti'' visualisations. In these case, overplotting all information on a single SED is very confusing (see Fig. \ref{fig:ex2_singlesed}). While, we keep this figure for its pedagogic value, we prefer to focus on the split version of SED to interpret the fitting (see Fig. \ref{fig:ex2_splitsed}). On the top panel (best-fit visualisation), the procedure proposes a really good fitting. When incorporating the uncertainties on the parameters, in the bottom panel (spaghetti visualisation), we see that the superposition of the different components explain the degeneracies observed in Fig. \ref{fig:ex2_triangle}. The large variation of the temperature allows for a large variation of the normalisation and, in turns, a different setup and relative contribution of each component. This is a perfect illustration that (i) a simple value of minimum $\chi^2$ fit is limited as not providing a fair view of the fitting, (ii) extra-information on the relative contribution of the different components is necessary to better constrain the fit (iii) even a sophisticated fitting approach does not preclude to high-quality data and a careful examination of the source. 

\subsection{Example \#3: real dataset}
\label{sec:ex3}

We finally run \mrmoose\ on a published dataset to ensure the code is able to recover previous results. We take the example from \cite{Gilli2014} as photometry and parameter results are reported in the publication and allow for a direct comparison. Briefly, the data, a combination of ALMA and {\it Herschel} observations with upper limits and detections, are used to fit a modified black body component (reported in \S~\ref{sec:models}), in order to access the dust properties of a $z$=4.75 quasar. We run the code in two configurations $a$ and $b$: one with only the detected filters and a second adding the upper limits to the fit. For both configurations we set $\beta$=2 and $\nu_0$=1.5\,10$^{12}$ Hz as in the original publication. We list the files in the Appendix and run the procedure twice. We focus only on recovering the temperature parameter as the other parameters are derived either from the best fit temperature or directly from observed fluxes. 

\begin{figure}
\includegraphics[width=0.5\textwidth]{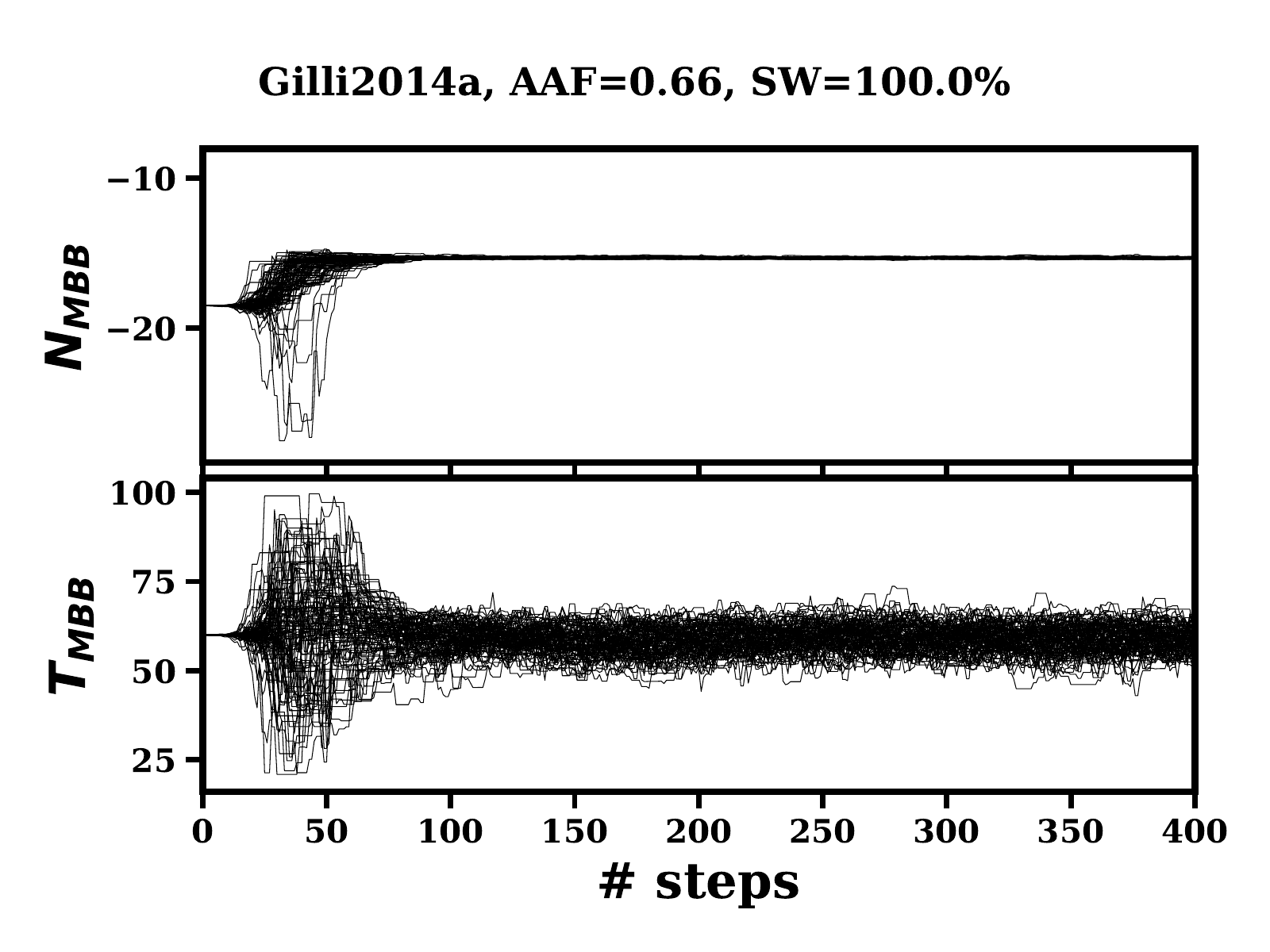} \\
\includegraphics[width=0.5\textwidth]{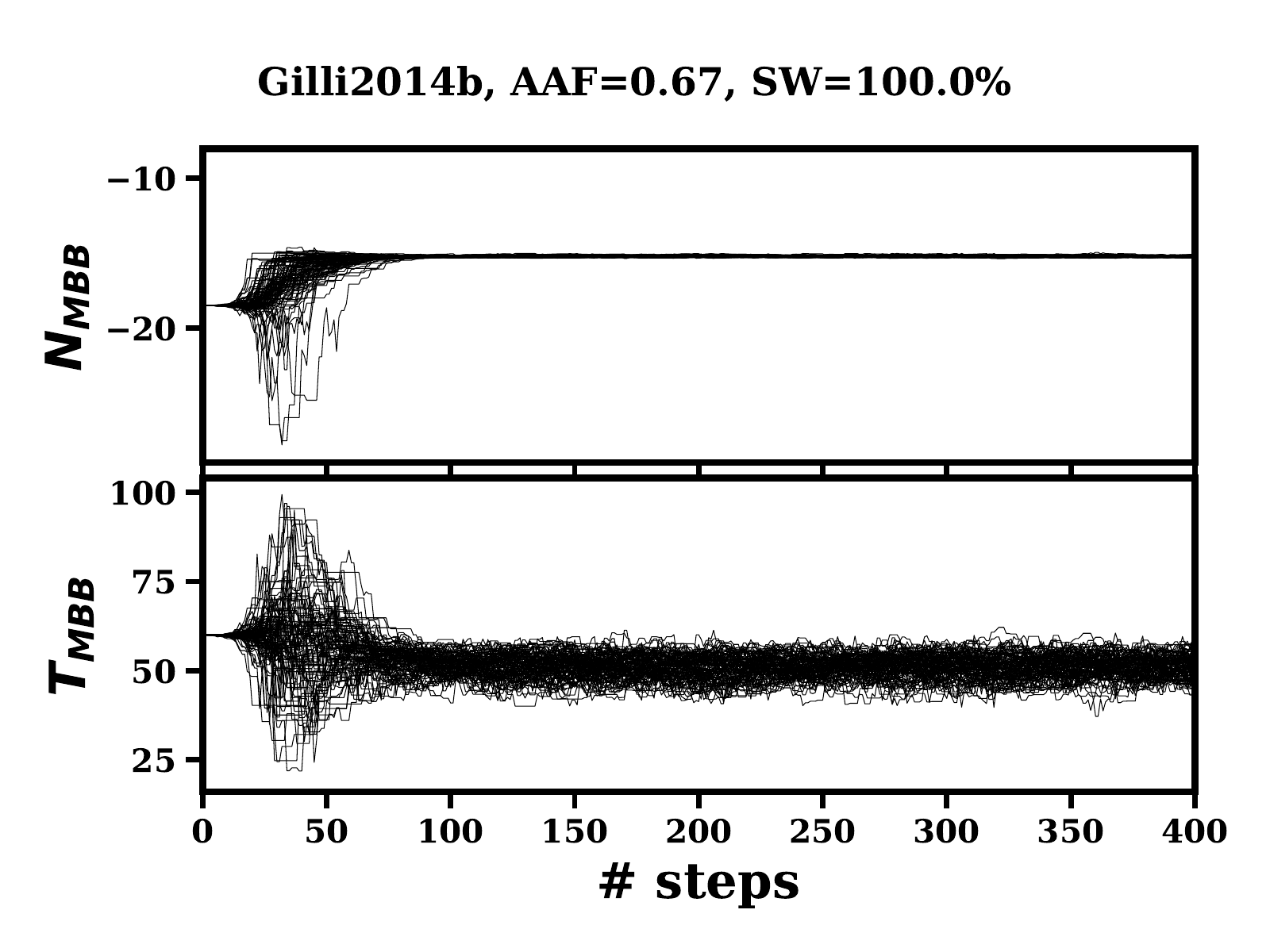}
\caption{MC-Chains for example \#3. Note the convergence after roughly 100 steps. {\it Top}: detections only{\it (a)}. {\it Bottom:} detections and upper limits{\it (b)}. }
\label{fig:ex3_mcchains}
\end{figure}

\begin{figure}
\includegraphics[width=0.5\textwidth]{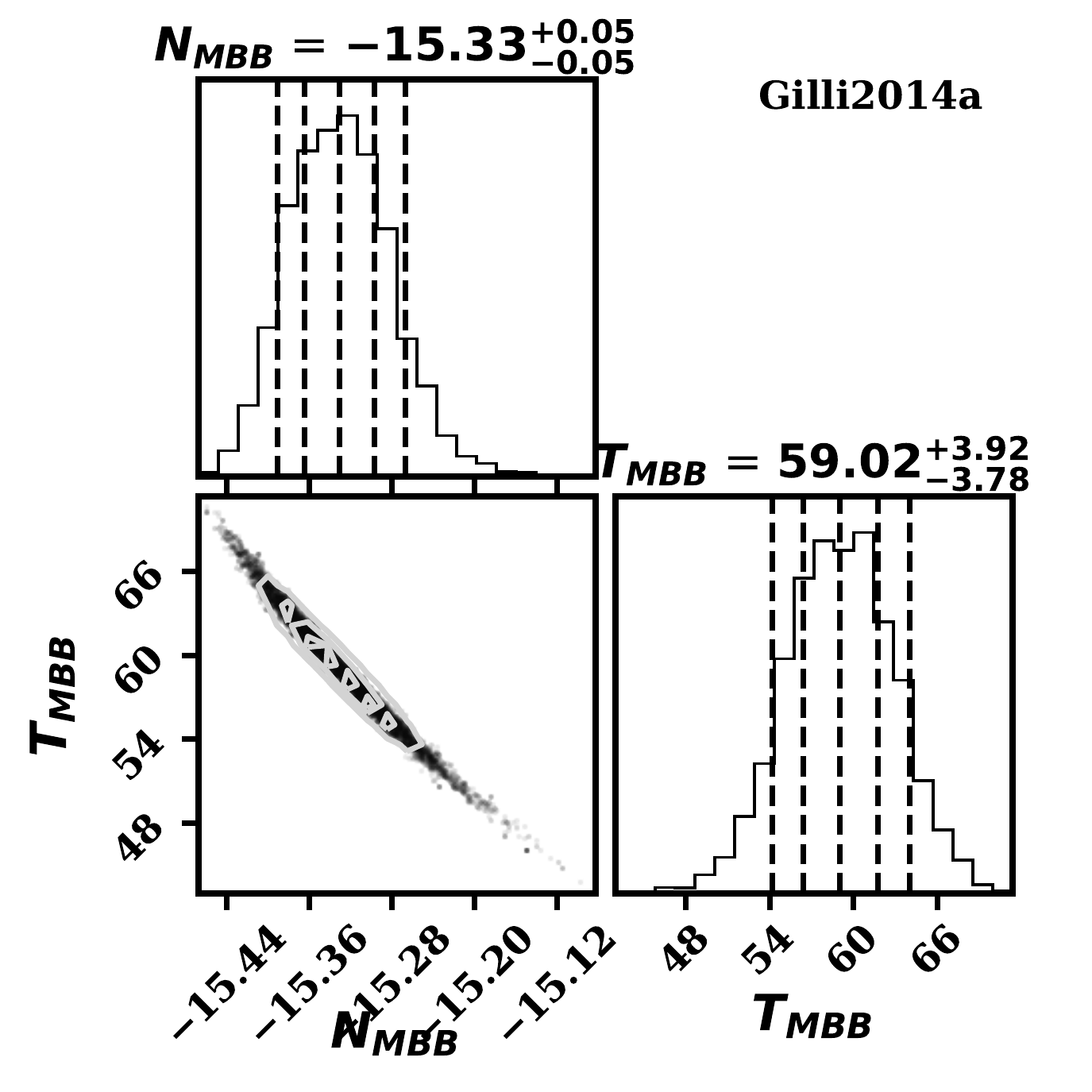} \\
\includegraphics[width=0.5\textwidth]{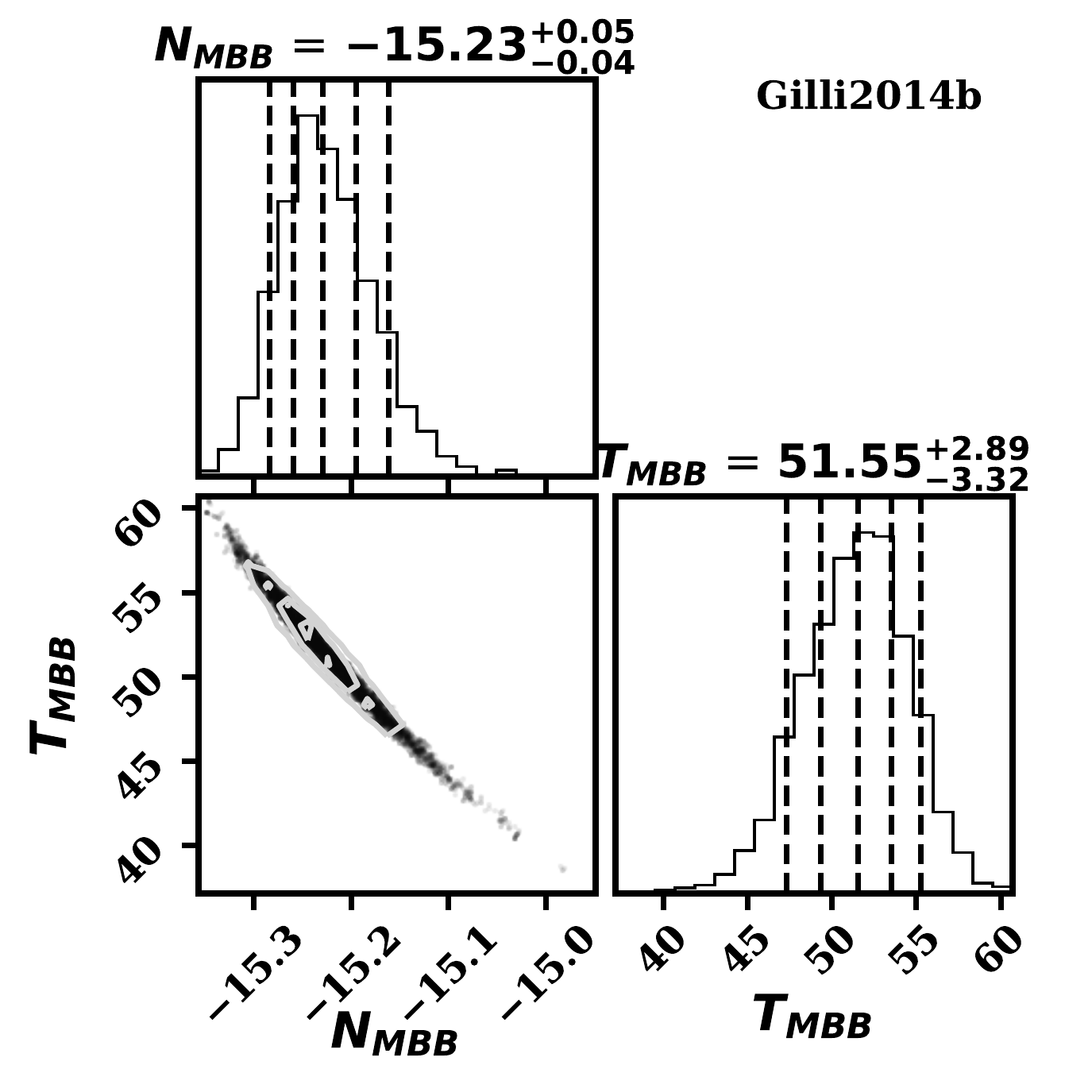}
\caption{Triangle diagram for example \#3. {\it Top}: detections only {\it (a)}. {\it Bottom:} detections and upper limits {\it (b)}. The diagonal represents the marginalised probability distribution. The vertical dashed line correspond to the 10\%, 25\%, 50\%, 75\% and 90\% percentiles of the distribution. The 50\% percentile parameter along the 25\% and 75\% percentile values are reported in Table \ref{tab:ex3_param}. The other diagram represents the 2D projection of the parameter space on the corresponding parameters on the abscissa and ordinate axes. The contours represents the 75\%, 50\% and 25\% of the maximum parameter value .}
\label{fig:ex3_triangle}
\end{figure}

\begin{figure*}
\centering
\begin{tabular}{cc}
\includegraphics[width=0.48\textwidth]{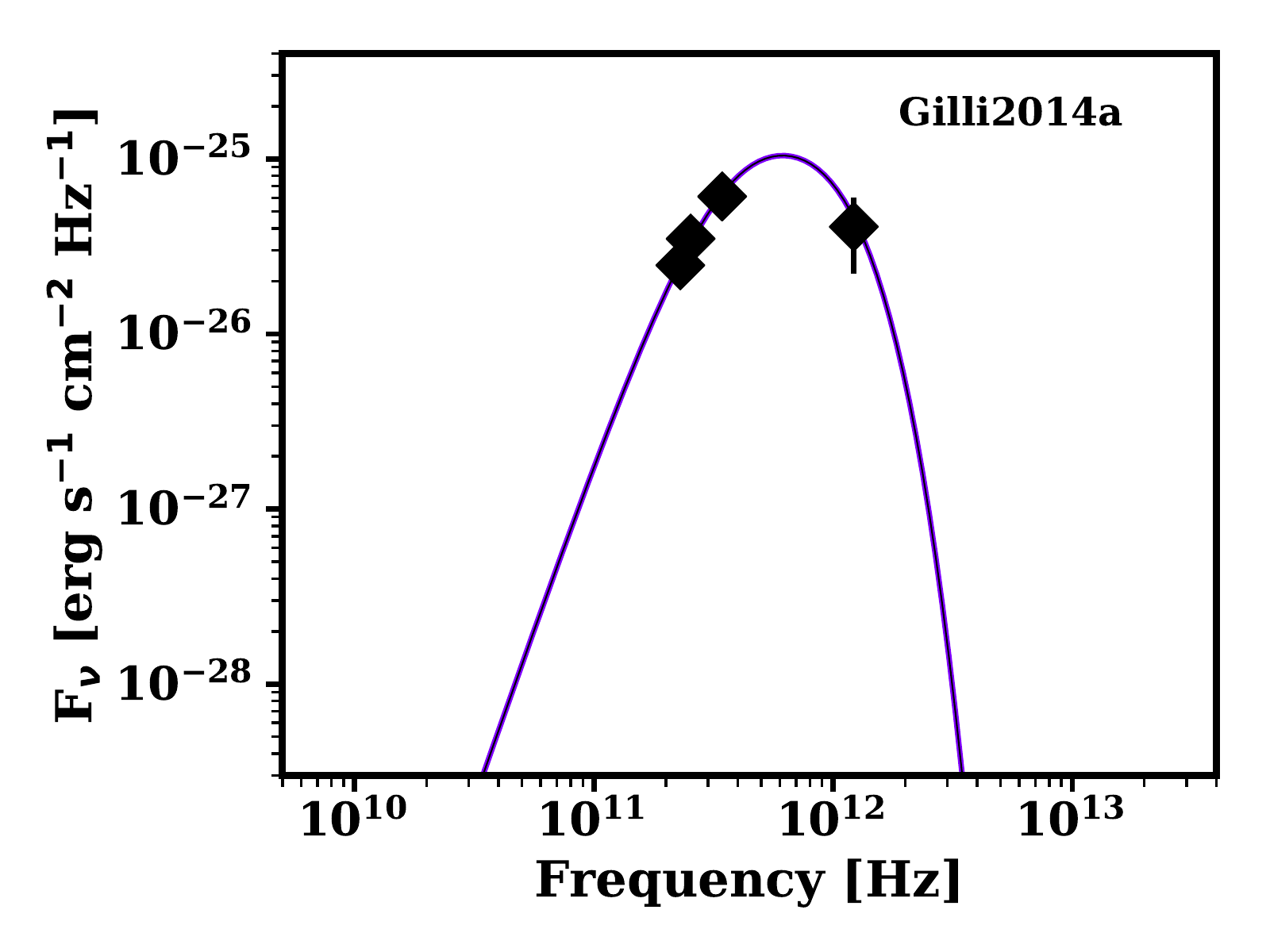} 
& \includegraphics[width=0.48\textwidth]{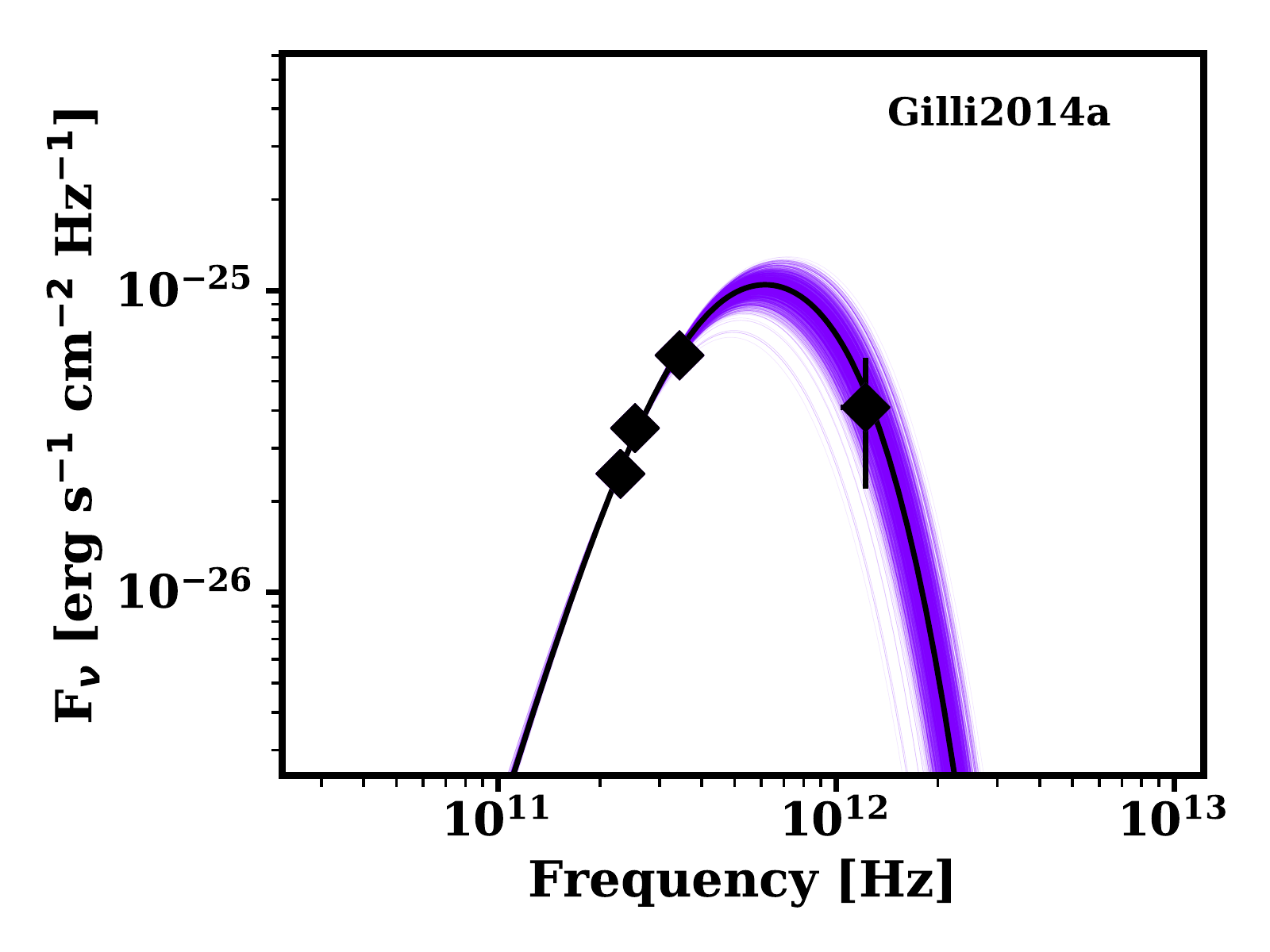} \\
\includegraphics[width=0.48\textwidth]{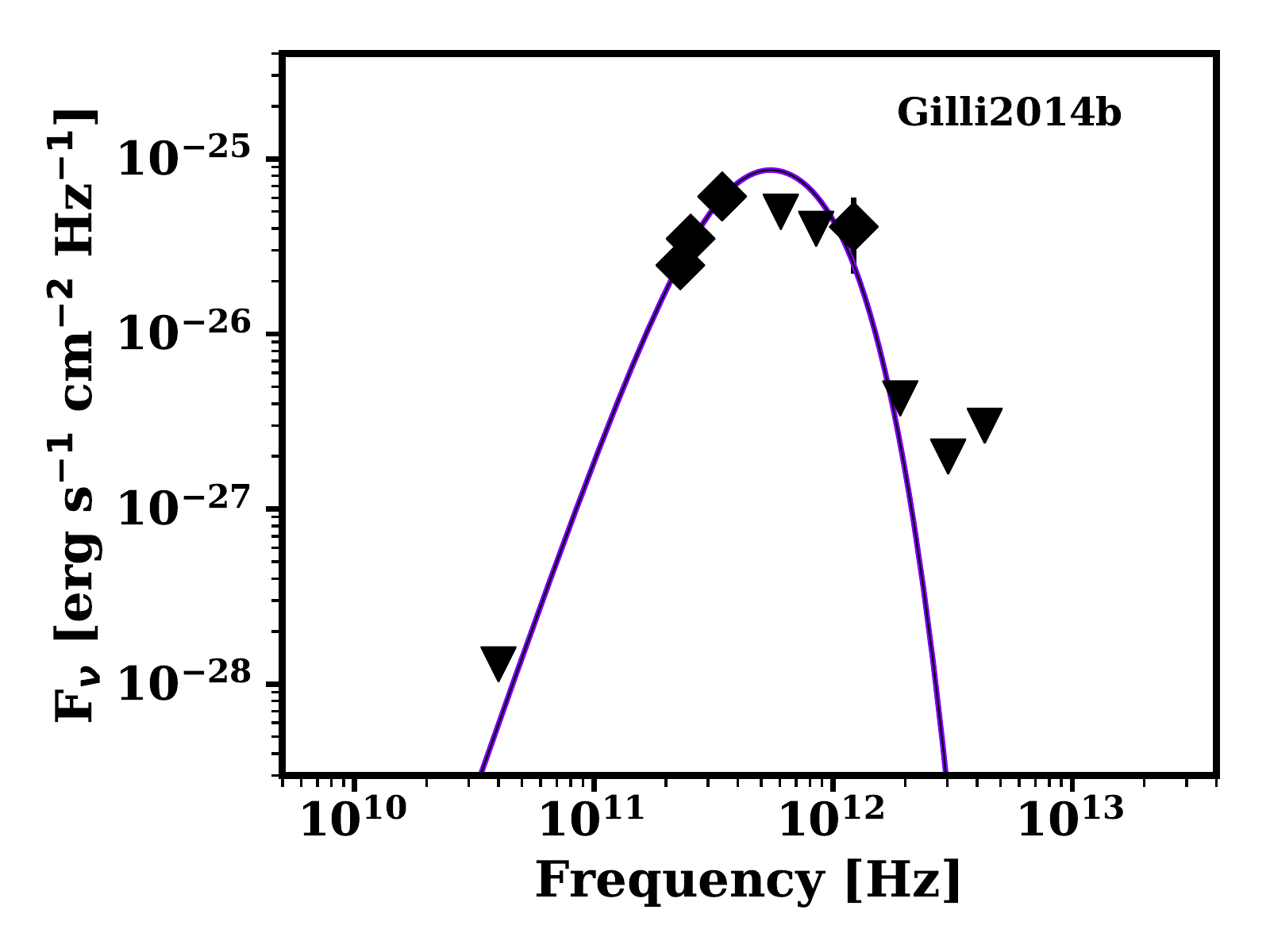} 
& \includegraphics[width=0.48\textwidth]{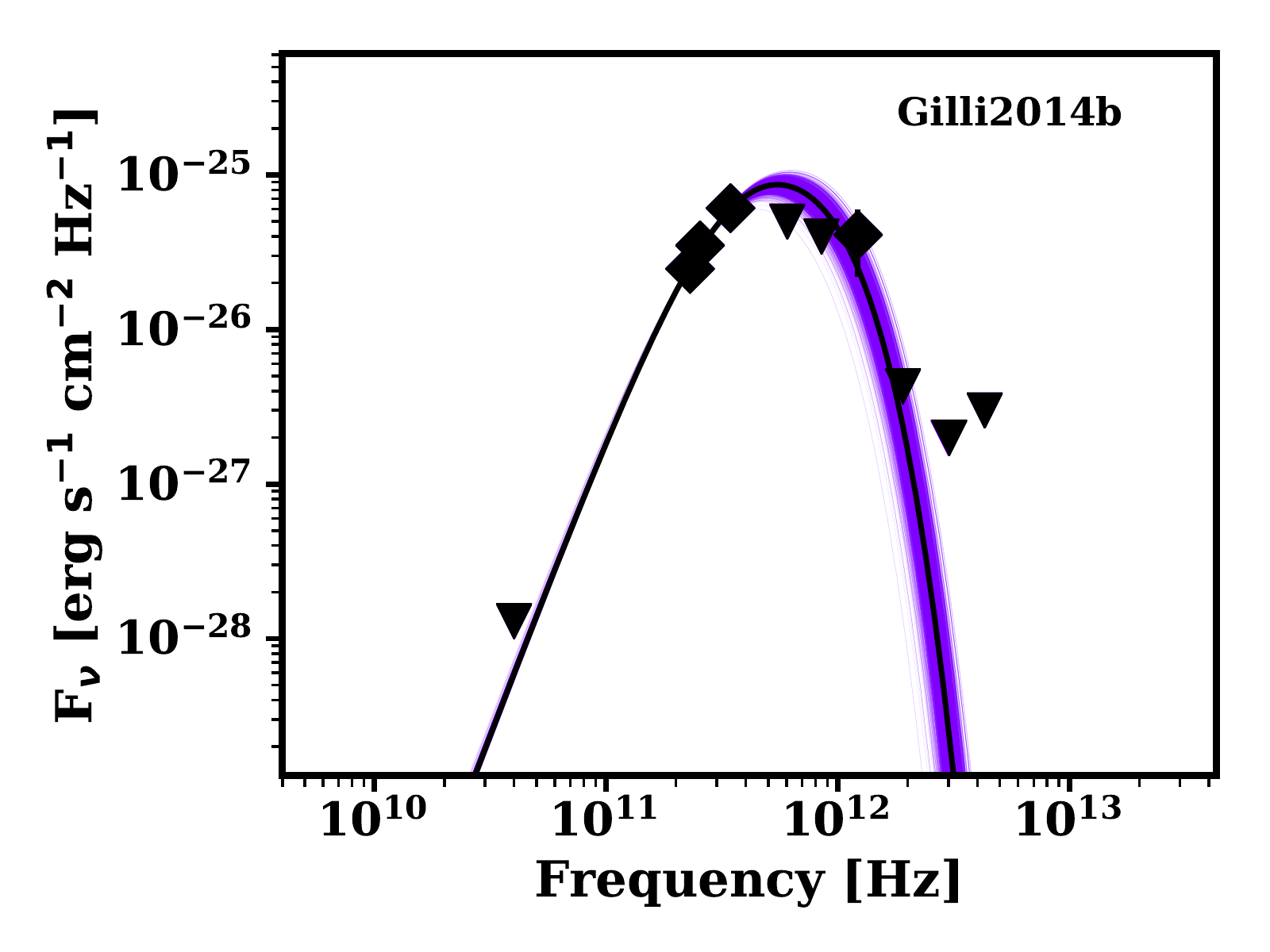} \\
\end{tabular}
\caption{SEDs for example \#3. {\it Top}: detections only{\it (a)}. {\it Bottom:} detections and upper limits{\it (b)}. {\it Left:} Best fit plot. {\it Right:} Spaghetti plot. The diamonds are the detections and the triangles are the 1$\sigma$ upper limits. Note that the fitting can converge above the upper limits to a certain extent.}
\label{fig:ex3_sed}
\end{figure*}

For the case of detection only {\it (a)}, the temperature is recovered within the uncertainties of the quoted value in the original paper (see Table \ref{tab:ex3_param}, Fig. \ref{fig:ex3_mcchains}, Fig. \ref{fig:ex3_triangle} and Fig. \ref{fig:ex3_sed}). Note also the asymmetrical uncertainties when having the information from parameter space exploration. The uncertainties provided by \mrmoose\ are different, referring to the 25\% and 75\% of the distribution when the value quoted in \cite{Gilli2014} is likely the standard 1$\sigma$ symmetric uncertainty assuming a normal distribution. 

When including upper limits {\it (b)}, the result differs with a slightly lower temperature and uncertainties. This is probably due to the inclusion of upper limits as continuous data with decreasing probability as defined in Eq. \ref{eq:chi2} (note also how the model is able to converge above the upper limits, which are represented as 1$\sigma$ in Fig. \ref{fig:ex3_sed}, bottom right rather than the 3$\sigma$ in Fig. 2, \cite{Gilli2014}. This is a good demonstration of how constraining upper limits can add valuable information when performing SED fitting. As mentioned in the original paper, we would like to stress that the uncertainties reported here are mainly fit uncertainties: they only represent the combined user knowledge on the uncertainties given in the data file.

For a more general comment, these statistical uncertainties do not incorporate any systematic uncertainties (which can dominate in certain cases). This is a very common problem in SED fitting, the final uncertainties obtained are limited to the systematic uncertainties associated to the assumed model to represent the data or from the data themselves. We add this word of caution for user non-familiar with SED fitting as these effects are often more subtle (but can be larger than the statistical uncertainties!) and usually hidden in the model choice and/or data. In this particular case, only a full comparison between a range of models (e.g. the common problem of IMF in stellar population fitting) and a complete data processing uncertainties (e.g. bad calibration or sky subtraction) assessment can answer this particular question.

\begin{table} \centering
\caption{Components and parameters of example \#3. $^*$ value in logarithmic scale. The uncertainties quoted are the 25 and 75 percentiles. See \S~\ref{sec:ex3} for the details about configurations {\it a} and {\it b}.}
\label{tab:ex3_param}
\begin{tabular}{ccccc}
\hline
Name & parameter & Gilli2014+ & config {\it a}  & config {\it b} \\ 
\hline \hline 
\multirow{2}{*}{MBB} & $N$ & n/a & -15.33$^{+0.05}_{-0.04}$ & -15.22$^{+0.05}_{-0.04}$ \\ [0.2ex]
                                  & $T$[K] & 58.5$^{+5.8}_{-5.8}$ & 59.14$^{+3.61}_{-4.06}$  &  51.08$^{+3.01}_{-3.23}$ \\
\hline
\end{tabular}
\end{table}

\section{Known limitations and future developments}
\label{sec:limitations}

\begin{table}
\caption{Execution time for different configurations of \mrmoose. All times reported are approximate time when executed on a relatively recent laptop (here for a laptop with a (3.1GHz Intel core i7 and 16GB DDR3 memory)).}
\label{tab:exec}
\centering
\begin{tabular}{ccccc}
\hline
Examples & \nwalkers & \nsteps & \# of parameters & Time \\
\hline \hline
\# 1 & 200 & 200 & 2 & $\sim$2\,min \\
\# 2 & 400 & 3500 & 10 & $\sim$24\,h \\
\# 3a & 200 & 400 & 2 & $\sim$5\,min \\
\# 3b & 200 & 400 & 2 & $\sim$10\,min \\
\hline
\end{tabular}
\end{table}

At the publication of this paper, the code will be available online on the GitHub platform at the following url: \url{https://github.com/gdrouart/MrMoose}, under a GPLv3 license (allowing re-use and modifications, see {\tt License.txt} in \mrmoose\ folder). While \mrmoose\ is under continuous development and will see periodic upgrades, the code also knows some limitations. We list some of the most important ones and refer the reader to the README for a more exhaustive list of all known issues/limitations.
\begin{itemize}
\item The execution time is highly dependent on the complexity of the configuration: the most parameters/components/arrangements, the longer the time for the fitting to converge. Typical execution times are from couple of minutes to several hours (see Table~\ref{tab:exec}) for a reasonably recent machine.  More particularly, it assumes that the user defined models are as efficient as possible because these models are called at each iteration for each walker and is therefore the bottleneck of the code. Large number of walkers and steps can lead to significantly larger files, slowing the process. Even if \mrmoose\ is designed to run in one step, we recommend to split the run in several steps, first generate the images only to check your data-file, hence a second run to perform the fit and a third to generate the ``.pdf'' files (by changing the keywords values in the setting file).
\item Choosing initial values close to the ``true'' values is important, otherwise the code will take a long time to converge (if converging at all). This choice of initial guess is particularly important when including upper limits. By design, when the parameter values produce a model above an upper limit, the likelihood is very quickly set to infinity due to the error function (see \S~\ref{sec:design}). Therefore a slight underestimation of parameters is preferable, particularly on any normalisation parameters (to keep the second term of the $\chi^2$ in Eq. \ref{eq:chi2} close to null at the beginning). We remind that the initialisation of the walkers is made following the {\tt emcee} recommendation: a Gaussian ``ball'' of walkers centred on the median value of the interval parameters.
\item The code provides two levels of parallelisation (multi-core processing) for the calculation of the likelihood: the parallelisation on one source and the parallelisation on a sample of sources. The former tends to deteriorate the performance, most likely due to the large overheads when the likelihood is relatively simple (see {\tt emcee} documentation). The second is only applicable on a sample of sources and therefore is much more efficient in reducing fitting time of an entire sample, using one core per source (\mrmoose\ is provided with the wrapper {\tt herd.py} to parallelise on sample). However, specific care should be taken to avoid running out of memory as the number of positions to store scales with the number of steps and walkers (and the number of sources to fit simultaneously).
\end{itemize}

We list here some of the future features, already planned: 
\begin{itemize}
\item allowing for a wider range of priors: in the current version, only uniform priors are incorporated. We plan to add a variety of priors in order to increase the fidelity of prior knowledge into the fitting procedure. This will also solve the initial parameter value limitations mentioned previously. 
\item a migration to Python 3 is planned (when upgraded, all further developments will be applied to the Python 3 in priority)
\item implementing the use of discrete models such as sophisticated AGN models or galaxy evolution models. This step requires the implementation of a methodology where the samplers can explore a (partially) discrete parameter space. Several solutions are considered (SED interpolation, brute force, branch-and-bound methods, nesting, etc). Its implementation is beyond the scope of this paper, it will be the object of a major update of \mrmoose.
\end{itemize}

\section{Conclusion}
\label{sec:conclusion}

We presented a new fitting tool allowing the user to fit SEDs for a variety of source configuration from relatively simple to complex source configuration in a bayesian framework using the {\tt emcee} package. The code, highly customisable, allows a wide range of  configurations between data and models, with the aim to stay user-friendly. The code consistently handles upper limits along with detections, each data point being calculated through the filter database, also managed and defined by the user. The large flexibility on creating models/data/filters makes this code virtually adaptable to any case, independently of wavelengths, instruments/telescopes combinations or science cases, providing a robust tool to interpret increasingly complex multi-wavelength datasets. In particular, the code appears ideally suited (but not limited) to fit a combination of {\it Spitzer}, {\it Herschel}, ALMA and JVLA dataset, and/or in the presence of constraining upper limits in the data even in the case of a single unresolved source.

\section*{Acknowledgements}
GD would like to thanks Steven Murray for insightful discussions on Bayesian inference and the Curtin PyClub for useful programming tips. We also thank Matt Lehnert, Carlos De Breuck, Joel Vernet and Nick Seymour for advices and help in the conception of this code. GD acknowledges the ESO Scientific Visitor Programme in supporting the development of this code. 

\bibliographystyle{mnras} 
\bibliography{Library}

\begin{thebibliography}{}
\makeatletter
\relax
\def\mn@urlcharsother{\let\do\@makeother \do\$\do\&\do\#\do\^\do\_\do\%\do\~}
\def\mn@doi{\begingroup\mn@urlcharsother \@ifnextchar [ {\mn@doi@}
  {\mn@doi@[]}}
\def\mn@doi@[#1]#2{\def\@tempa{#1}\ifx\@tempa\@empty \href
  {http://dx.doi.org/#2} {doi:#2}\else \href {http://dx.doi.org/#2} {#1}\fi
  \endgroup}
\def\mn@eprint#1#2{\mn@eprint@#1:#2::\@nil}
\def\mn@eprint@arXiv#1{\href {http://arxiv.org/abs/#1} {{\tt arXiv:#1}}}
\def\mn@eprint@dblp#1{\href {http://dblp.uni-trier.de/rec/bibtex/#1.xml}
  {dblp:#1}}
\def\mn@eprint@#1:#2:#3:#4\@nil{\def\@tempa {#1}\def\@tempb {#2}\def\@tempc
  {#3}\ifx \@tempc \@empty \let \@tempc \@tempb \let \@tempb \@tempa \fi \ifx
  \@tempb \@empty \def\@tempb {arXiv}\fi \@ifundefined
  {mn@eprint@\@tempb}{\@tempb:\@tempc}{\expandafter \expandafter \csname
  mn@eprint@\@tempb\endcsname \expandafter{\@tempc}}}

\bibitem[\protect\citeauthoryear{{Asensio Ramos} \& {Ramos Almeida}}{{Asensio
  Ramos} \& {Ramos Almeida}}{2009}]{AsensioRamos2009}
{Asensio Ramos} A.,  {Ramos Almeida} C.,  2009, \mn@doi [\apj]
  {10.1088/0004-637X/696/2/2075}, \href
  {http://adsabs.harvard.edu/abs/2009ApJ...696.2075A} {696, 2075}

\bibitem[\protect\citeauthoryear{{Astropy Collaboration} et~al.,}{{Astropy
  Collaboration} et~al.}{2013}]{astropy2013}
{Astropy Collaboration} et~al., 2013, \mn@doi [\aap]
  {10.1051/0004-6361/201322068}, \href
  {http://adsabs.harvard.edu/abs/2013A%26A...558A..33A} {558, A33}

\bibitem[\protect\citeauthoryear{Bowley}{Bowley}{1920}]{Bowley1920}
Bowley A. L. (Arthur~Lyon) S.,  1920, An elementary manual of statistics, 3rd
  ed. edn.
P. S. King \& Son, \url {http://ci.nii.ac.jp/ncid/BB1069398X}

\bibitem[\protect\citeauthoryear{{Bruzual} \& {Charlot}}{{Bruzual} \&
  {Charlot}}{2003}]{Bruzual2003}
{Bruzual} G.,  {Charlot} S.,  2003, \mn@doi [\mnras]
  {10.1046/j.1365-8711.2003.06897.x}, \href
  {http://cdsads.u-strasbg.fr/abs/2003MNRAS.344.1000B} {344, 1000}

\bibitem[\protect\citeauthoryear{{Calistro Rivera}, {Lusso}, {Hennawi}  \&
  {Hogg}}{{Calistro Rivera} et~al.}{2016}]{CalistroRivera2016}
{Calistro Rivera} G.,  {Lusso} E.,  {Hennawi} J.~F.,   {Hogg} D.~W.,  2016,
  \mn@doi [\apj] {10.3847/1538-4357/833/1/98}, \href
  {http://adsabs.harvard.edu/abs/2016ApJ...833...98C} {833, 98}

\bibitem[\protect\citeauthoryear{{Conroy}}{{Conroy}}{2013}]{Conroy2013}
{Conroy} C.,  2013, \mn@doi [\araa] {10.1146/annurev-astro-082812-141017},
  \href {http://adsabs.harvard.edu/abs/2013ARA%26A..51..393C} {51, 393}

\bibitem[\protect\citeauthoryear{{Drouart} \& {et al.}}{{Drouart} \& {et
  al.}}{prep}]{Drouart2017}
{Drouart} G.,  {et al.} in prep, \mnras

\bibitem[\protect\citeauthoryear{{Falkendal} et~al.,}{{Falkendal}
  et~al.}{2018}]{Falkendal2017}
{Falkendal} T.,  et~al., sub., 2018, \aap

\bibitem[\protect\citeauthoryear{{Fanaroff} \& {Riley}}{{Fanaroff} \&
  {Riley}}{1974}]{Fanaroff1974}
{Fanaroff} B.~L.,  {Riley} J.~M.,  1974, \mnras, \href
  {http://adsabs.harvard.edu/abs/1974MNRAS.167P..31F} {167, 31P}

\bibitem[\protect\citeauthoryear{{Fioc} \& {Rocca-Volmerange}}{{Fioc} \&
  {Rocca-Volmerange}}{1997}]{Fioc1997}
{Fioc} M.,  {Rocca-Volmerange} B.,  1997, \aap, \href
  {http://cdsads.u-strasbg.fr/abs/1997A%26A...326..950F} {326, 950}

\bibitem[\protect\citeauthoryear{{Fiorucci} \& {Munari}}{{Fiorucci} \&
  {Munari}}{2003}]{Fiorucci2003}
{Fiorucci} M.,  {Munari} U.,  2003, \mn@doi [\aap]
  {10.1051/0004-6361:20030075}, \href
  {http://adsabs.harvard.edu/abs/2003A%26A...401..781F} {401, 781}

\bibitem[\protect\citeauthoryear{{Foreman-Mackey}, {Hogg}, {Lang}  \&
  {Goodman}}{{Foreman-Mackey} et~al.}{2013}]{emcee2013}
{Foreman-Mackey} D.,  {Hogg} D.~W.,  {Lang} D.,   {Goodman} J.,  2013, \mn@doi
  [\pasp] {10.1086/670067}, \href
  {http://adsabs.harvard.edu/abs/2013PASP..125..306F} {125, 306}

\bibitem[\protect\citeauthoryear{{Fritz}, {Franceschini}  \&
  {Hatziminaoglou}}{{Fritz} et~al.}{2006}]{Fritz2006}
{Fritz} J.,  {Franceschini} A.,   {Hatziminaoglou} E.,  2006, \mn@doi [\mnras]
  {10.1111/j.1365-2966.2006.09866.x}, \href
  {http://cdsads.u-strasbg.fr/abs/2006MNRAS.366..767F} {366, 767}

\bibitem[\protect\citeauthoryear{{Gilli} et~al.,}{{Gilli}
  et~al.}{2014}]{Gilli2014}
{Gilli} R.,  et~al., 2014, \mn@doi [\aap] {10.1051/0004-6361/201322892}, \href
  {http://adsabs.harvard.edu/abs/2014A%26A...562A..67G} {562, A67}

\bibitem[\protect\citeauthoryear{{Goodman} \& {Weare}}{{Goodman} \&
  {Weare}}{2010}]{Goodman2010}
{Goodman} J.,  {Weare} J.,  2010, \mn@doi [Communications in Applied
  Mathematics and Computational Science, Vol.~5, No.~1, p.~65-80, 2010]
  {10.2140/camcos.2010.5.65}, \href
  {http://adsabs.harvard.edu/abs/2010CAMCS...5...65G} {5, 65}

\bibitem[\protect\citeauthoryear{{Groves}, {Dopita}  \& {Sutherland}}{{Groves}
  et~al.}{2004}]{Groves2004}
{Groves} B.~A.,  {Dopita} M.~A.,   {Sutherland} R.~S.,  2004, \mn@doi [\apjs]
  {10.1086/421113}, \href {http://adsabs.harvard.edu/abs/2004ApJS..153....9G}
  {153, 9}

\bibitem[\protect\citeauthoryear{{Gullberg} et~al.,}{{Gullberg}
  et~al.}{2016}]{Gullberg2016}
{Gullberg} B.,  et~al., 2016, \mn@doi [\aap] {10.1051/0004-6361/201526858},
  \href {http://adsabs.harvard.edu/abs/2016A%26A...586A.124G} {586, A124}

\bibitem[\protect\citeauthoryear{{Harwood}, {Hardcastle}, {Croston}  \&
  {Goodger}}{{Harwood} et~al.}{2013}]{Harwood2013}
{Harwood} J.~J.,  {Hardcastle} M.~J.,  {Croston} J.~H.,   {Goodger} J.~L.,
  2013, \mn@doi [\mnras] {10.1093/mnras/stt1526}, \href
  {http://adsabs.harvard.edu/abs/2013MNRAS.435.3353H} {435, 3353}

\bibitem[\protect\citeauthoryear{{H{\"o}nig}, {Beckert}, {Ohnaka}  \&
  {Weigelt}}{{H{\"o}nig} et~al.}{2006}]{Honig2006}
{H{\"o}nig} S.~F.,  {Beckert} T.,  {Ohnaka} K.,   {Weigelt} G.,  2006, \mn@doi
  [\aap] {10.1051/0004-6361:20054622}, \href
  {http://cdsads.u-strasbg.fr/abs/2006A%26A...452..459H} {452, 459}

\bibitem[\protect\citeauthoryear{{Leitherer} et~al.,}{{Leitherer}
  et~al.}{1999}]{Leitherer1999}
{Leitherer} C.,  et~al., 1999, \mn@doi [\apjs] {10.1086/313233}, \href
  {http://cdsads.u-strasbg.fr/abs/1999ApJS..123....3L} {123, 3}

\bibitem[\protect\citeauthoryear{{MacKenzie}, {Scott}  \&
  {Swinbank}}{{MacKenzie} et~al.}{2016}]{MacKenzie2016}
{MacKenzie} T.~P.,  {Scott} D.,   {Swinbank} M.,  2016, \mn@doi [\mnras]
  {10.1093/mnras/stw1890}, \href
  {http://adsabs.harvard.edu/abs/2016MNRAS.463...10M} {463, 10}

\bibitem[\protect\citeauthoryear{{Maraston}}{{Maraston}}{1998}]{Maraston1998}
{Maraston} C.,  1998, \mn@doi [\mnras] {10.1046/j.1365-8711.1998.01947.x},
  \href {http://adsabs.harvard.edu/abs/1998MNRAS.300..872M} {300, 872}

\bibitem[\protect\citeauthoryear{{Nenkova}, {Sirocky}, {Ivezi{\'c}}  \&
  {Elitzur}}{{Nenkova} et~al.}{2008}]{Nenkova2008a}
{Nenkova} M.,  {Sirocky} M.~M.,  {Ivezi{\'c}} {\v Z}.,   {Elitzur} M.,  2008,
  \mn@doi [\apj] {10.1086/590482}, \href
  {http://cdsads.u-strasbg.fr/abs/2008ApJ...685..147N} {685, 147}

\bibitem[\protect\citeauthoryear{{Noll}, {Burgarella}, {Giovannoli}, {Buat},
  {Marcillac}  \& {Mu{\~n}oz-Mateos}}{{Noll} et~al.}{2009}]{Noll2009}
{Noll} S.,  {Burgarella} D.,  {Giovannoli} E.,  {Buat} V.,  {Marcillac} D.,
  {Mu{\~n}oz-Mateos} J.~C.,  2009, \mn@doi [\aap]
  {10.1051/0004-6361/200912497}, \href
  {http://adsabs.harvard.edu/abs/2009A%26A...507.1793N} {507, 1793}

\bibitem[\protect\citeauthoryear{{Pier} \& {Krolik}}{{Pier} \&
  {Krolik}}{1992}]{Pier1992}
{Pier} E.~A.,  {Krolik} J.~H.,  1992, \mn@doi [\apj] {10.1086/172042}, \href
  {http://cdsads.u-strasbg.fr/abs/1992ApJ...401...99P} {401, 99}

\bibitem[\protect\citeauthoryear{{Pilbratt} et~al.,}{{Pilbratt}
  et~al.}{2010}]{Pilbratt2010}
{Pilbratt} G.~L.,  et~al., 2010, \mn@doi [\aap] {10.1051/0004-6361/201014759},
  \href {http://adsabs.harvard.edu/abs/2010A%26A...518L...1P} {518, L1}

\bibitem[\protect\citeauthoryear{{Sawicki}}{{Sawicki}}{2012}]{Sawicki2012}
{Sawicki} M.,  2012, \mn@doi [\pasp] {10.1086/668636}, \href
  {http://adsabs.harvard.edu/abs/2012PASP..124.1208S} {124, 1208}

\bibitem[\protect\citeauthoryear{{Sharma}}{{Sharma}}{2017}]{Sharma2017}
{Sharma} S.,  2017, preprint, \href
  {http://adsabs.harvard.edu/abs/2017arXiv170601629S} {} (\mn@eprint {arXiv}
  {1706.01629})

\bibitem[\protect\citeauthoryear{{Siebenmorgen}, {Heymann}  \&
  {Efstathiou}}{{Siebenmorgen} et~al.}{2015}]{Siebenmorgen2015}
{Siebenmorgen} R.,  {Heymann} F.,   {Efstathiou} A.,  2015, \mn@doi [\aap]
  {10.1051/0004-6361/201526034}, \href
  {http://adsabs.harvard.edu/abs/2015A%26A...583A.120S} {583, A120}

\bibitem[\protect\citeauthoryear{{Siringo} et~al.,}{{Siringo}
  et~al.}{2009}]{Siringo2009}
{Siringo} G.,  et~al., 2009, \mn@doi [\aap] {10.1051/0004-6361/200811454},
  \href {http://adsabs.harvard.edu/abs/2009A%26A...497..945S} {497, 945}

\bibitem[\protect\citeauthoryear{{Stalevski}, {Fritz}, {Baes}, {Nakos}  \&
  {Popovi{\'c}}}{{Stalevski} et~al.}{2012}]{Stalevski2012}
{Stalevski} M.,  {Fritz} J.,  {Baes} M.,  {Nakos} T.,   {Popovi{\'c}} L.~{\v
  C}.,  2012, \mn@doi [\mnras] {10.1111/j.1365-2966.2011.19775.x}, \href
  {http://adsabs.harvard.edu/abs/2012MNRAS.420.2756S} {420, 2756}

\bibitem[\protect\citeauthoryear{{Tinsley}}{{Tinsley}}{1968}]{Tinsley1968}
{Tinsley} B.~M.,  1968, \mn@doi [\apj] {10.1086/149455}, \href
  {http://adsabs.harvard.edu/abs/1968ApJ...151..547T} {151, 547}

\bibitem[\protect\citeauthoryear{{Wells}, {Greisen}  \& {Harten}}{{Wells}
  et~al.}{1981}]{Wells1981}
{Wells} D.~C.,  {Greisen} E.~W.,   {Harten} R.~H.,  1981, \aaps, \href
  {http://adsabs.harvard.edu/abs/1981A%26AS...44..363W} {44, 363}

\bibitem[\protect\citeauthoryear{{Werner} et~al.,}{{Werner}
  et~al.}{2004}]{Werner2004}
{Werner} M.~W.,  et~al., 2004, \mn@doi [\apjs] {10.1086/422992}, \href
  {http://adsabs.harvard.edu/abs/2004ApJS..154....1W} {154, 1}

\bibitem[\protect\citeauthoryear{{da Cunha}, {Charlot}  \& {Elbaz}}{{da Cunha}
  et~al.}{2008}]{daCunha2008}
{da Cunha} E.,  {Charlot} S.,   {Elbaz} D.,  2008, \mn@doi [\mnras]
  {10.1111/j.1365-2966.2008.13535.x}, \href
  {http://adsabs.harvard.edu/abs/2008MNRAS.388.1595D} {388, 1595}

\makeatother
\end{thebibliography}

\clearpage

\appendix
\section*{List of packages used in \mrmoose\ and references}

The list of packages, their version and their respective reference.
\begin{itemize}
\item astropy v2.0rc1, http://www.astropy.org, \citet{astropy2013}
\item corner v2.0.1, https://pypi.python.org/pypi/corner
\item emcee v2.2.1, http://dan.iel.fm/emcee/current/\#, \citet{emcee2013}
\item guppy v0.1.10, https://pypi.python.org/pypi/guppy/
\item matplotlib v1.5.1, http://matplotlib.org/
\item numpy v1.9.1, http://www.numpy.org/
\item pathos v0.2.0, https://pypi.python.org/pypi/pathos/0.2.0
\item pycallgraph v1.0.1, http://pycallgraph.slowchop.com/en/master/\#, 
\item PyYAML v3.12, http://pyyaml.org/
\item tqdm v4.8.4, https://github.com/tqdm/tqdm
\item scipy v0.14.0, https://www.scipy.org/
\end{itemize}

\section*{Input files for both examples.} 

\begin{lstlisting}[caption={Example \#1 of model file. Note the latex notation of the parameter (used in the generation of the plots)}, label={lst:mod1}]{}
sync_law  2 
$N$       -25  -15 
$\alpha$  -2.0  0.0 
\end{lstlisting}

\begin{lstlisting}[caption={Example \#1 of the fit file.}, label={lst:fit1}]{}
source_file: data/fake_source_ex1.dat 
model_file: models/fake_source_ex1.mod
all_same_redshift: True 
redshift: [0.0,]
nwalkers: 200 
nsteps: 200 
nsteps_cut: 180 
percentiles: [10., 25., 50., 75., 90.] 
skip_imaging: False
skip_fit: False
skip_MCChains: False
skip_triangle: False
skip_SED: False 
unit_obs: 'Hz' 
unit_flux: 'Jy' 
\end{lstlisting}

\begin{lstlisting}[caption={Example \#2 of model file. Note the latex notation of the parameter (used in the generation of the plots)}, label={lst:mod2}]{}
AGN_law  2 
$N_{AGN}$   -38  -28 
$\alpha_{AGN}$ -4.0  0.0 
BB_law  2 
$N_{BB1}$   -28  -18 
$T_1$  10  80 
BB_law  2 
$N_{BB2}$   -28  -18 
$T_2$  10  80 
sync_law  2 
$N_{s1}$   -28  -18 
$\alpha_{s1}$ -2.0  0.0 
sync_law  2 
$N_{s2}$   -28  -18 
$\alpha_{s2}$ -2.0  0.0
\end{lstlisting}

\begin{lstlisting}[caption={Example \#2 of the fit file.}, label={lst:fit2}]{}
source_file: data/fake_source_ex6.dat 
model_file: models/fake_source_ex6.mod
all_same_redshift: True 
redshift: [2.0, 2.0, 2.0, 2.0, 2.0]
nwalkers: 400 
nsteps: 3500 
nsteps_cut: 3450 
percentiles: [10., 25., 50., 75., 90.] 
skip_imaging: False
skip_fit: False
skip_MCChains: False
skip_triangle: False
skip_SED: False 
unit_obs: 'Hz' 
unit_flux: 'Jy' 
\end{lstlisting}

\begin{lstlisting}[caption={Example \#3 of model file. Note the latex notation of the parameter (used in the generation of the plots)}, label={lst:mod3}]{}
MBB_law  2 
$N$   -28  -9 
$T$   20  100 
\end{lstlisting}

\begin{lstlisting}[caption={Example \#3 of the fit file.}, label={lst:fit1}]{}
source_file: data/Gilli2014a.dat 
model_file: models/MBB.mod
all_same_redshift: True 
redshift: [4.75,]
nwalkers: 200 
nsteps: 400 
nsteps_cut: 350 
percentiles: [10., 25., 50., 75., 90.] 
skip_imaging: False
skip_fit: False
skip_MCChains: False
skip_triangle: False
skip_SED: False 
unit_obs: 'Hz' 
unit_flux: 'Jy' 
\end{lstlisting}

\begin{landscape}
\begin{lstlisting}[float=*,caption={Example \#1 of data file. The wavelength references are reported in Hz and the flux are reported in Jy (10$^{-23}$ erg s$^{-1}$ cm$^{-2}$ Hz$^{-1}$).}, label={lst:data1}, linewidth= 24cm,basicstyle=\small]{}
# filter        RA              Dec        resolution  lambda0  det_type  flux   flux_error  arrangement  component   component_number 
74MHz(VLA)      12h00m00s       -40d00m00s       12.0 7.377042e+07 d     1.188135e-07 9.036851e-09 1          sync       0          
408MHz          12h00m00s       -40d00m00s       12.0 4.078365e+08 d     2.526585e-08 1.634557e-09 1          sync       0          
1.4GHz          12h00m00s       -40d00m00s       12.0 1.399439e+09 d     7.190462e-09 4.763989e-10 1          sync       0          
4.85GHz         12h00m00s       -40d00m00s       12.0 4.848056e+09 d     2.188714e-09 1.375087e-10 1          sync       0          
8.4GHz          12h00m00s       -40d00m00s       12.0 8.396633e+09 d     1.161870e-09 7.939377e-11 1          sync       0,        
\end{lstlisting}
 
\begin{lstlisting}[float=*,caption={Example \#2 of data file. The wavelength references are reported in Hz and the flux are reported in Jy (10$^{-23}$ erg s$^{-1}$ cm$^{-2}$ Hz$^{-1}$).}, label={lst:data2}, linewidth= 24cm,basicstyle=\small]{}
# filter        RA              Dec        resolution  lambda0  det_type  flux   flux_error  arrangement  component   component_number                                                      
VLA_L           12h00m00s       -40d00m00s       20.0 1.400000e+09 d     6.642489e-08 1.310263e-08 4          'total sync' 3,4                                                              
VLA_C           12h00m00.1s     -39d59m59s        1.0 4.710000e+09 d     1.726653e-08 3.875161e-09 6          'E sync'   4                                                                  
VLA_C           11h59m59.95s    -40d00m01s        1.0 4.710000e+09 d     1.994406e-08 3.875161e-09 5          'W sync'   3                                                                  
VLA_X           12h00m00.1s     -39d59m59s        0.5 8.210000e+09 d     1.115211e-08 2.226486e-09 6          'E sync'   4                                                                  
VLA_X           11h59m59.95s    -40d00m01s        0.5 8.210000e+09 d     9.245703e-09 2.226486e-09 5          'W sync'   3                                                                  
ATCA_47         12h00m00s       -40d00m00s       10.0 4.799800e+10 d     2.015150e-09 4.231155e-10 4          'total sync' 3,4                                                              
ALMA_3          12h00m00s       -40d00m00s        3.0 1.022500e+11 d     1.701354e-09 3.677694e-10 7          'all sources' 0,1,2,3,4                                                       
ALMA_6          12h00m00.1s     -39d59m59s        0.3 2.452500e+11 d     5.485048e-09 9.451120e-10 3          'E sync/host' 2,4                                                             
ALMA_6_nr1      11h59m59.95s    -40d00m00.5s      0.3 2.291500e+11 d     4.714464e-09 8.537784e-10 2          'W sync/comp' 1,3                                                             
laboca_870      12h00m00s       -40d00m00s       15.0 3.440046e+11 u     9.107478e-09 9.107478e-09 1          'Dust/AGN' 0,1,2                                                              
spire_500       12h00m00s       -40d00m00s       35.0 6.048278e+11 u     1.595681e-08 1.595681e-08 1          'Dust/AGN' 0,1,2                                                              
spire_350       12h00m00s       -40d00m00s       25.0 8.498941e+11 u     1.553297e-08 1.553297e-08 1          'Dust/AGN' 0,1,2                                                              
spire_250       12h00m00s       -40d00m00s       17.0 1.221085e+12 d     1.666475e-08 3.975734e-09 1          'Dust/AGN' 0,1,2                                                              
pacs_160        12h00m00s       -40d00m00s        5.0 1.909476e+12 d     1.166623e-08 2.640865e-09 1          'Dust/AGN' 0,1,2                                                              
pacs_70         12h00m00s       -40d00m00s        4.0 4.304211e+12 d     1.416861e-09 3.018548e-10 1          'Dust/AGN' 0,1,2                                                              
mips_24         12h00m00s       -40d00m00s        7.0 1.283529e+13 d     2.471502e-10 4.419324e-11 1          'Dust/AGN' 0,1,2                                                              
irs_16          12h00m00s       -40d00m00s        5.0 1.946903e+13 d     1.054598e-10 2.169550e-11 1          'Dust/AGN' 0,1,2                                                              
irac_4          12h00m00s       -40d00m00s        5.0 3.909498e+13 d     3.153092e-11 6.182625e-12 1          'Dust/AGN' 0,1,2
\end{lstlisting}

\begin{lstlisting}[float=*,caption={Example \#3 of data file. The wavelength references are reported in Hz and the flux are reported in Jy (10$^{-23}$ erg s$^{-1}$ cm$^{-2}$ Hz$^{-1}$). Note that in configuration $a$, only the detection are used (see \S~\ref{sec:ex3}).}, label={lst:data1}, linewidth= 24cm,basicstyle=\small]{}
# filter        RA              Dec        resolution  lambda0  det_type  flux   flux_error  arrangement  component   component_number 
pacs_70         03h32m29.35s    -27d56m19.6s      4.0 4.304211e+12 u      0.3e-3    0.3e-3    1           'Dust'      0
pacs_100        03h32m29.35s    -27d56m19.6s      4.0 3.000000e+12 u      0.2e-3    0.2e-3    1           'Dust'      0
pacs_160        03h32m29.35s    -27d56m19.6s      5.0 1.909476e+12 u      0.43e-3   0.43e-3   1           'Dust'      0
spire_250       03h32m29.35s    -27d56m19.6s     17.0 1.221085e+12 d      4.1e-3    1.9e-3    1           'Dust'      0
spire_350       03h32m29.35s    -27d56m19.6s     25.0 8.498941e+11 u      4.0e-3    4.0e-3    1           'Dust'      0
spire_500       03h32m29.35s    -27d56m19.6s     35.0 6.048278e+11 u      5.0e-3    5.0e-3    1           'Dust'      0      
ALMA_B7_344     03h32m29.35s    -27d56m19.6s      1.5 3.440000e+11 d      6.1e-3    0.5e-3    1           'Dust'      0
ALMA_B6_254     03h32m29.35s    -27d56m19.6s      1.5 2.540000e+11 d      3.5e-3    0.1e-3    1           'Dust'      0
ALMA_B6_230     03h32m29.35s    -27d56m19.6s      0.8 2.299000e+11 d      2.47e-3   0.07e-3   1           'Dust'      0
ATCA_40GHz      03h32m29.35s    -27d56m19.6s     10.0 4.000000e+10 u      0.013e-3  0.013e-3  1           'Dust'      0,
\end{lstlisting}
\end{landscape}

%%%%%%%%%%%%%%%%%%%%%%%%%%%%%%%%%%%%%%%%%%%%%%%%%%

% Don't change these lines
\bsp	% typesetting comment
\label{lastpage}
\end{document}